**Zipf's Law for All the Natural Cities around the World**

Bin Jiang[1], Junjun Yin[2], and Qingling Liu[1]

[1]Department of Technology and Built Environment, Division of Geomatics
University of Gävle, SE-801 76 Gävle, Sweden
Email: bin.jiang@hig.se

[2]Department of Geography and Geographic Information Science
University of Illinois at Urbana and Champaign, USA
Email: jyn@illinois.edu

*(Draft: December 2013, Revision: May, August, September, and October 2014)*

**Abstract**
Two fundamental issues surrounding research on Zipf's law regarding city sizes are whether and why this law holds. This paper does not deal with the latter issue with respect to why, and instead investigates whether Zipf's law holds in a global setting, thus involving all cities around the world. Unlike previous studies, which have mainly relied on conventional census data such as populations, and census-bureau-imposed definitions of cities, we adopt naturally (in terms of data speaks for itself) delineated cities, or natural cities, to be more precise, in order to examine Zipf's law. We find that Zipf's law holds remarkably well for all natural cities at the global level, and remains almost valid at the continental level except for Africa at certain time instants. We further examine the law at the country level, and note that Zipf's law is violated from country to country or from time to time. This violation is mainly due to our limitations; we are limited to individual countries, or to a static view on city-size distributions. The central argument of this paper is that Zipf's law is universal, and we therefore must use the correct scope in order to observe it. We further find Zipf's law applied to city numbers; the number of cities in the first largest country is twice as many as that in the second largest country, three times as many as that in the third largest country, and so on. These findings have profound implications for big data and the science of cities.

**1. Introduction**
A widely observed regularity for cities around the world is that city sizes are inversely proportional to their ranks. Put simply, by ranking all the cities (or to be more precise, human settlements) of a country, from the largest to the smallest according to their populations, one can note that the largest city is twice as big as the second largest, and three times as big as the third largest, and so on. This regularity, known as Zipf's law (named after the linguist George Kingsley Zipf (1949)), was first discovered by the German physicist Felix Auerbach (1913). In the literature, Zipf's law is often used to refer to a power law in general, with an exponent of between 0 and 2. However, in the context of this paper, we stick to Zipf's law with an exponent of one ($\pm 0.1$). Remarkably, this law has been known to hold for at least 100 hundred years (Gabaix 1999, Li 2002, Mitzenmacher 2003, Newman 2005). Despite its ubiquity, some researchers (e.g., Soo 2005) tend to be skeptical of Zipf's law for all cities, and even Gabaix (1999), who seems to be no skeptical of the law, admit its validity for the largest cities in some occasions. This skepticism surrounds two basic questions: (1) does the law apply to all cities (or just large cities) within a country? (2) does the law apply to cities in all countries (or just large countries)? In the literature, whether Zipf's law holds varies from one country to another; it is valid for large cities, while small cities better fit alternative others such as a lognormal distribution. Zipf's law is not universal, as many others have claimed. Therefore, the literature provides us with a contradictory picture about Zipf's law: it has been claimed that it is universal, but from time to time the law has been violated.

The two questions above are not always legitimate from a scientific point of view, in particular in the era of big data. If the number of cities in a country is too small (for example, < 6 like Singapore), Zipf's law would not hold; this is because it indicates a statistical regularity, which needs a sufficient big sample. Conventionally cities within individual countries have been examined in order to verify Zipf's law. This is understandable, since all cities in a country are usually considered to be an interconnected or interdependent whole. However, this interconnectedness is not always at play; for example, the impacts of primate cities (Jefferson 1939) go beyond their country borders. Given the circumstances, we should not be constrained by the cities within individual countries, but rather the cities that are truly considered as an interconnected whole. Given these backgrounds, a legitimate question would be: does Zipf's law apply to all cities around the world (or only to large cities around the world)? This paper is primarily motivated by this question. Our central argument is that Zipf's law is universal, and the reason we fail to observe Zipf's law is due to our limitations; we are limited to census data, individual countries, or a static view of city-size distributions. Therefore, it is important to use the correct perspective and scope when observing Zipf's law. The country scope is not always legitimate, given that (1) some countries have too few cities to reveal the statistical regularity, and (2) the impacts of some cities, so called global or world cities such as New York, London, and Tokyo (Saskia 1991) go beyond their country borders. It is difficult to determine a whole of cities, but there is little doubt that all cities around the world constitute an interconnected whole, just as all people on the planet constitute a socially connected whole. We therefore claim that Zipf's law applies to a whole of cities, rather than an arbitrary set of cities (see Section 5 for further discussion on this).

This paper examines all natural cities extracted from satellite imagery; more specifically, it considers three night-time images taken during a 19-year period for the entire world. All the images were inter-calibrated, so that the pixels' values are comparable from one year to another. Natural cities are naturally and objectively delineated human settlements using a single-pixel-value cutoff (see Section 2 for technical details). Instead of using a conventional least squares estimate, we utilized the most robust maximum likelihood estimate for power law detection for all of the natural cities respectively at the global, continental, and country levels. The novelty of this paper can be seen from the following three aspects: (1) we studied Zipf's law and verified its universality in a global setting involving all natural cities as a whole around the world; (2) we abandoned the use of census-bureau-imposed cities, and instead adopted natural cities from night-time imagery of the world; and (3) we found that city numbers among individual countries follow Zipf's law; the number of cities in the first largest country is twice as many as that in the second largest country, three times as many as that in the third largest country, and so on. Overall, in this study we provided a new perspective on the dispute surrounding Zipf's law, and examined the law in the context of big data.

The remainder of this paper is structured as follows. Section 2 presents the concept of natural cities, and how they can be extracted from the night-time imagery. Section 3 introduces Zipf's law and its equivalents (the Pareto distribution and power law) by a working example, and briefly presents methods on how to detect Zipf's law, in particular using the most robust maximum likelihood estimates. Section 4 outlines results on the verification of Zipf's law at the global, continental and country levels, as well as detailing other related results. The implications of the study are further discussed in Section 5. Finally, Section 6 draws a conclusion and points to future work.

**2. Natural cities extracted from night-time imagery**
There is a large body of literature within the field of remote sensing on how to extract cities or equivalently human settlements from satellite imagery (e.g., Yang 2011). Indeed, remote sensing imagery provides a powerful means by which to delimit cities in terms of their extents and locations, but there is no guarantee that cities can be automatically extracted from the imagery (Weber 2001). The underlying reason for this is not related to the methods per se, but rather to the very definition of cities. The conventional definition of cities is a product of census, which literally means "to estimate" populations for taxation purposes. According to the US Census Bureau, "cities" could refer to incorporated places, urban areas, and/or metropolitan districts with certain population thresholds. On the other hand, remote sensing imagery aims not to estimate, but to accurately and precisely record or measure, to the best of the sensor's ability, what is on the Earth's surface. It is no wonder that an estimated entity often does not match an accurately recorded one. In addition, cities are continuously changing. For example, an authority may have decided, legally and administratively, that a certain piece of land should become part of a city, but remote sensing imagery, not to mention censuses, may not have recorded this. Given these arguments, conventional definitions of cities imposed from the top down by authorities remain valid for many scientific studies, but should not be the only choice. Jiang and his co-workers (Jiang and Jia 2011, Jiang and Liu 2012) suggested the notion of "natural cities" as an alternative to conventional definitions of cities based on populations.

Natural cities refer to human settlements, or human activities in general on the Earth's surface, that are objectively and naturally delineated from massive geographic information of various kinds, and based on the head/tail division rule. The key to the definition is the head/tail division rule (Jiang and Liu 2012): for any variable, if its values follow a heavy tailed distribution such as a power law, lognormal and exponential distributions, then all values can be split around the average into two unbalanced parts: a minority above the average as the "head," and a majority below the average as the "tail." This rule was formulated when Jiang and Liu (2012) extracted millions of city (and field) blocks within a country, and categorized those that were below the average size as "natural cities." Surprisingly, the extracted natural cities are fair reflections of conventionally defined cities based on populations (or real cities), at a collective level, in the three largest European countries: France, Germany and the UK. Interestingly, the head/tail division rule can be applied to values in the head recursively, leading to a new classification scheme called head/tail breaks (Jiang 2013). In this paper, natural cities are defined and delineated from night-time satellite imagery.

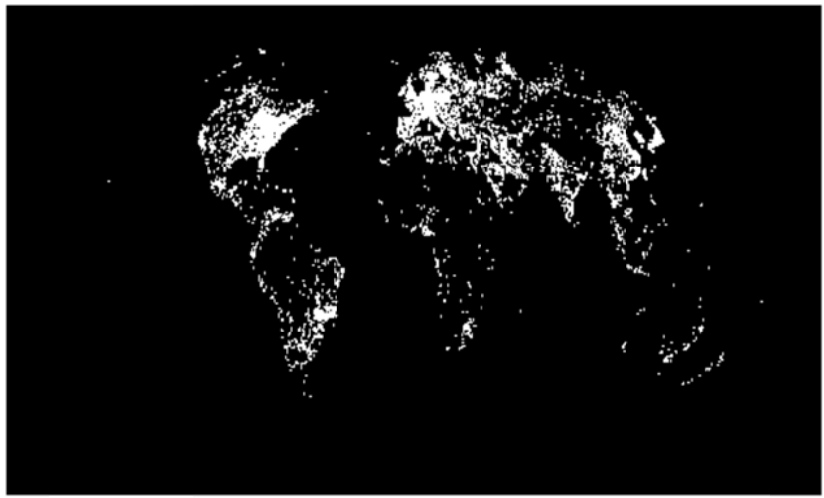

Figure 1: Night-time image captured in 2010 for the Mollweide equal area projection

The night-time imagery data was obtained from the National Oceanic and Atmospheric Administration's National Geophysical Data Center (http://ngdc.noaa.gov/eog/dmsp/downloadV4composites.html). Each image contains 725,820,001 pixels, with values ranging from 0 (darkest) to 63 (lightest). For the time period 1992–2010 there are a total of 31 images, captured by six satellites. The resolution of the night-time images is 30 arc-seconds, for approximately 1 kilometer at the equator. In this study, we chose three images from 1992, 2001 and 2010, respectively. The three images were firstly calibrated based on the techniques developed by Elvidge et al. (2013), in order for the pixel values to be comparable across different years. To accurately measure natural cities, we adopted the Mollweide equal area projection (Figure 1).

We utilize the US night-time image as a working example to illustrate how to determine a threshold for deriving natural cities. The same example was used to illustrate the computation of ht-index (Jiang and Yin 2014). The US night-time image contains a total of 11,766,012 pixels, with light values or gray scales between 0 and 63. There are far more dark pixels than light ones, so we can adopt the head/tail breaks to derive the hierarchy of the image. Firstly, the average lightness of the 11,766,012 pixels is 7.5 (the first mean), which splits all pixels into two unbalanced parts: 3,091,666 pixels (26 percent) above the first mean in the head, and 8,674,346 pixels (74 percent) below the mean in the tail (see first row in Table 1). The average lightness of the 3,091,666 pixels in the head is 23.0 (the second mean), which again splits the head into two unbalanced parts: 1,043,922 pixels in the head (34 percent) and 2,047,744 pixels in the tail (66 percent) (see second row in Table 1). If we continue the same partition for the 1,043,922 pixels, the two parts would not be unbalanced, but well balanced (50 to 50) (see third row in Table 1). The second mean would be the meaningful cutoff from which to derive the natural cities; all connected pixels lighter than 23.0 are grouped as individual natural cities (Figure 2). Note that we merge all pixels (with values greater than the threshold) that are located in proximities into individual natural cities. In this regard, we adopted a very loosely defined neighborhood, i.e., two pixels are neighboring as long as they touch either at a corner or side. In fact, defining natural cities using the first and third mean would be too fat and too thin, but those deriving from the second mean are just right. It should be noted that head/tail breaks relies on the average or arithmetic mean to break things into the head and tail, rather than use the mean to characterize things. This makes head/tail breaks (Jiang 2013) a powerful classification scheme for data with a heavy tailed distribution.

Table 1: Statistics for the head/tail breaks for the US night-time imagery
(Note: count = number of pixels; light*count = sum of individual light*count at each light level; # = number)

| Light | Count | Light*Count | Mean | # in head | % in head | # in tail | % in tail |
|---|---|---|---|---|---|---|---|
| 0–63 | 11,766,012 | 88,424,914 | 7.5 | 3,091,666 | 26 | 8,674,346 | 74 |
| 8–63 | 3,091,666 | 71,030,197 | 23.0 | 1,043,922 | 34 | 2,047,744 | 66 |
| 23–63 | 1,043,922 | 46,200,290 | 44.3 | 524,230 | 50 | 519,692 | 50 |

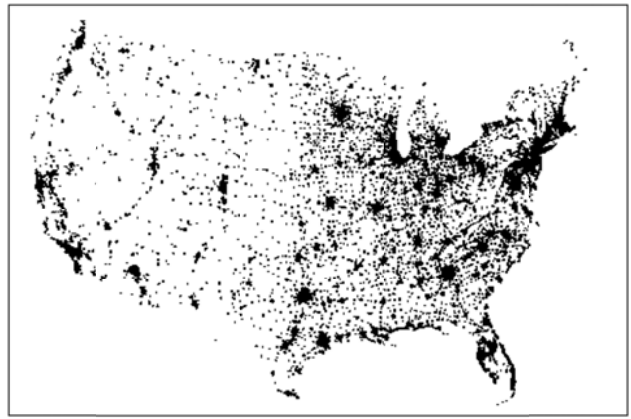

Figure 2: Natural cities derived from night-time imagery of the USA

The procedure illustrated above was applied to night-time images of the world for 1992, 2001, and 2010. Unlike previous studies that aimed to extract conventionally defined or imposed real cities (e.g., Tannier et al. 2011, Imhoff et al. 1997), with different thresholds for different countries, the natural cities are based on a single unique threshold. This study is not intended to setup a one-to-one relationship between the natural cities and real cities. As we will see later in the paper (Figure 5 in particular), the natural city in 2010 in the Shanghai region includes many real cities, so does the natural city in 2010 in the Nile river region. We believe that all of the cities in the world are an interconnected or interdependent whole, so we need a universal criterion in order to define urban agglomerations. This criterion is based on a collective decision regarding massive individual pixels, by averaging their values to obtain a meaningful cutoff by which to derive natural cities. Eventually, the sizes of the natural cities from the night-time images can be characterized by their physical extents rather than human populations. Before discussing the results, we shall first present methods on how to detect Zipf's law and power laws.

## 3. How to detect Zipf's law or power laws?

Zipf's law was initially formulated by $x = r^{-1}$, indicating an inverse power relationship between the city rank (r) and city size (x) (Zipf 1949). This inverse power relationship can be equivalently expressed as a cumulative distribution function (CDF), $(X \geq x) = x^{-1}$, where $Pr(X \geq x)$ can be considered equivalent to the number of cities greater than x. This CDF, also called Pareto distribution (initially discovered by the Italian economist Vilfredo Pareto), is concerned with the number of cities with a population greater than x. Both Zipf's law and Pareto distribution are more commonly seen as a probability distribution function (PDF), $pr(X = x) = x^{-2}$, referring to the number of cities whose population is exactly x, rather than greater than x. It should be noted that the power law exponent is two rather than one. For the sake of simplicity and accessibility to general audiences, as well as to geographers, we tried not to use more strict mathematical formulas than the three aforementioned; readers who are concerned with power law mathematics should refer to Newman (2005), Clauset et al. (2009), and references therein for more details. To make the paper self-contained, in this section we introduce Zipf's law, the Pareto distribution, and power law in general via a working example.

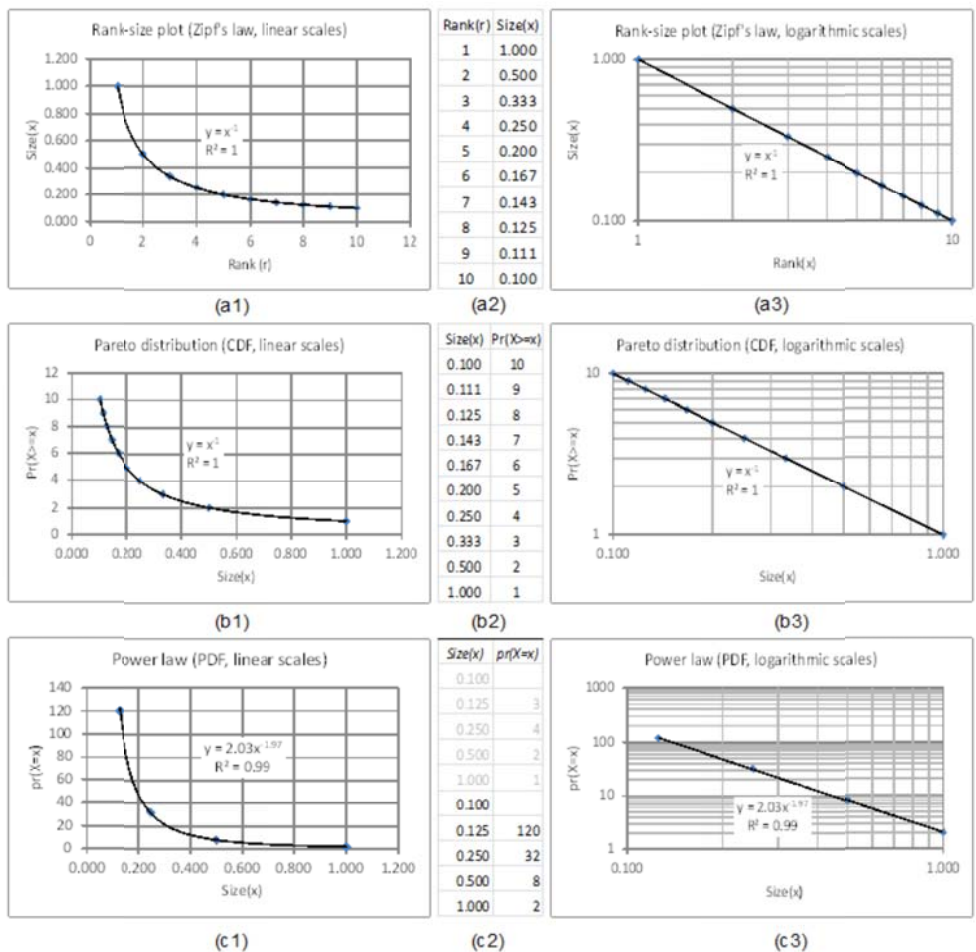

Figure 3: Illustration of Zipf's law, Pareto distributions and power laws
(Note: 10 synthetic city sizes: 1, 1/2, 1/3,…, and 1/10 are plotted according to Zipf's law, CDF, and PDF. The left column (Panels a1, b1, and c1) shows the plots in linear scales, while the right column (Panels a3, b3 and c3) shows the plots in logarithmic scales. The middle column (Panels a2, b2, and c2) displays the corresponding data of the plots. The PDF in the third row was obtained from the normalized frequency (120, 32, 8, and 2), rather than the initial frequency (3, 4, 2, and 1). For example, the frequency 120 was computed from 3/(0.125-0.10).)

We generated synthetic data representing 10 city sizes: x = 1, 1/2, 1/3,…, and 1/10, which follow Zipf's law exactly. The data was plotted according to Zipf's law, Pareto distribution (CDF), and power law (PDF), in terms of both linear and logarithmic scales, as shown in Figure 3. Obviously, the straight distribution lines emerged in the log-log plots (see Panels a3, b3, and c3 of Figure 3). Interested readers can type the synthetic data into an Excel sheet in order to duplicate these plots. A word of caution is in order for the PDF: there would be no straight distribution lines (or the tail end of the distribution would look messy for a big sample) if one used sizes (or "bins" to use the statistical term) with arithmetic progression, such as 0.1, 0.4, 0.7, and 1.0. We therefore used a geometric progression that increased by a factor of two: 0.10, 0.125, 0.25, 0.50, and 1.00. The PDF is in effect a probability density function, or the number of samples per unit width. Note that all of these plots were created using a least squares estimate; this is fine for the example, but would generate unreliable or biased results regarding power laws, as discussed in the literature (Adamic 2002, Newman 2005, Clauset et al. 2009). For this reason, in this study we adopted a more robust maximum likelihood estimate for the power law detection.

The maximum likelihood method is so far the most rigorous statistical examination of power laws. Accompanying this method is an index, p value, which is used to characterize the goodness of fit. Readers who are interested in the method and index should refer to Newman (2005) and Clauset et al. (2009), and the Matlab code (http://www.santafe.edu/~aaronc/powerlaws/) for technical details. In this study, we computed four parameters for each set of cities: the number of cities, the power law exponent alpha, the minimum city size, above which the cities exhibit a power law, and the goodness-of-fit index p value. In addition, for each set of cities, we verified Zipf's law (1) for all cities, (2) for those cities greater than the minimum, and (3) for large (greater than average) cities. This study is primarily concerned with whether city sizes exhibit Zipf's law, so we did not carry out an examination of any alternative distributions, such as lognormal, or a power law with an exponential cutoff.

## 4. Results and discussion

Based on the aforementioned procedure, we extracted about 30,000 natural cities in the world for each of the three years (1992, 2001, and 2010; see Figure 4 for an example). Note that for each of the three night-time images, one single threshold was used as a cutoff to derive the natural cities. We applied the head/tail breaks (Jiang 2013) to the three images, and obtained three thresholds, which are respectively 33, 29 and 31. Eventually, we used the average for the three thresholds (31) to derive the natural cities for the comparison purpose from one year to another. The extracted natural cities were then assessed in terms of whether they exhibit Zipf's law at three different levels: global, continent, and country. All of the natural cities are distributed among five continents, and over 230 countries or regions. We ran three separate tests for each set of cities: (1) for all of the natural cities, (2) for those greater than 10 square kilometers, and (3) for those greater than their average. The 10 square kilometers threshold is an estimate, above which Zipf's law holds remarkably, seen from the rank-size plots (panel b and c of Figure 4). The third test can be said for large cities, while the first and second tests for all cities.

We found that Zipf's law holds remarkably well at the global level for the 30,000 cities, and remains unchanged from one year to another. This is clearly reflected in Zipf's exponent of 1.0, and the high p value ($\geq$ 0.03) (Table 2). At the continental level, the power law generally still holds, with the exception of Africa in 2010, but the scaling exponent varies between 1.9 and 2.1, and with p values greater than 0.02; readers can cross-check the results in Table 2 (upper half). These results clearly indicate that Zipf's law remains valid, although it is not as striking at the continental level as at the global level. Obviously, Zipf's law is violated at the country level. In fact, no single country, including the USA, demonstrates Zipf's law as strikingly as it is exhibited at the global level, as the Zipf's exponent is round 1.0 at country level rather than exactly 1.0 at the global level. It should be noted that Zipf's exponent estimated using least squares would be higher than that estimated using maximum likelihood. For a comparison purpose, we deliberately chose the six countries in which Zipf's law was violated while using census-imposed population data (Benguigui and Blumenfeld-Lieberthal 2007, 2011, Cristelli et al. 2012). In fact, Zipf's law holds remarkably well for certain countries at certain times, as highlighted in Table 2. We examined 230 countries; 137 are presented in Appendix A, and the unlisted countries yield no valid statistical results due to the fact that the numbers of cities are fewer than six. As seen in Appendix A, a sufficient number of cities (for example, >200) is a precondition for holding Zipf's law, but not vice versa. It is important to note that violation of Zipf's law at the continental and country levels does not rule out the law's universality, because continents and countries do not provide the correct scope (see further discussion in the next section).

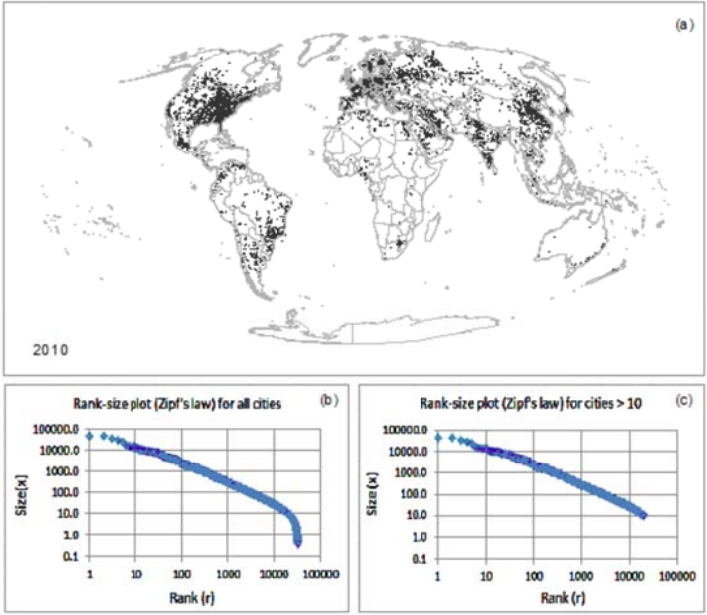

Figure 4: Natural cities and their rank-size plots: (a) 30,000 cities (black dots) for 2010, (b) rank-size plot for all 30,000 cities, and (c) rank-size plot for cities greater than 10 square kilometers

Table 2: Power law detection at the global, continent and country levels
(Note: At country level, only six countries are included in the table; refer to Appendix A for the complete list of 137 countries. # = number of cities, Zipf's exponent = alpha-1, xmin = minimum value for power law fit, p = goodness of fit; highlighted are those exhibiting Zipf's law.)

|  | 1992 | | | | 2001 | | | | 2010 | | | |
| --- | --- | --- | --- | --- | --- | --- | --- | --- | --- | --- | --- | --- |
|  | # | alpha | xmin | p | # | alpha | xmin | p | # | alpha | xmin | p |
| | | | | | WORLD | | | | | | | |
| All | 30278 | 2.0 | 23.9 | 0.03 | 27044 | 2.0 | 45.5 | 0.03 | 30747 | 2.0 | 56.2 | 0.03 |
| >10 | 16784 | 2.0 | 23.9 | 0.03 | 16000 | 2.0 | 45.5 | 0.02 | 17688 | 2.0 | 56.2 | 0.04 |
| >mean | 4064 | 2.0 | 52.6 | 0.04 | 3691 | 2.0 | 59.0 | 0.03 | 4007 | 2.0 | 64.6 | 0.03 |
| | | | | | AFRICA | | | | | | | |
| All | 1340 | 1.9 | 13.6 | 0.08 | 1336 | 1.9 | 19.9 | 0.07 | 1686 | 2.2 | 118.9 | 0.12 |
| >10 | 860 | 1.9 | 13.6 | 0.08 | 843 | 1.9 | 19.9 | 0.08 | 973 | 2.2 | 118.9 | 0.15 |
| >mean | 200 | 2.0 | 67.7 | 0.12 | 184 | 2.0 | 69.8 | 0.11 | 220 | 2.1 | 66.7 | 0.15 |
| | | | | | ASIA PACIFIC | | | | | | | |
| All | 9316 | 2.0 | 23.9 | 0.08 | 8321 | 2.0 | 50.8 | 0.08 | 10521 | 2.1 | 131.5 | 0.08 |
| >10 | 5157 | 2.0 | 23.9 | 0.07 | 5011 | 2.0 | 50.8 | 0.08 | 6280 | 2.1 | 131.5 | 0.07 |
| >mean | 1409 | 2.0 | 48.9 | 0.10 | 1256 | 2.0 | 56.4 | 0.06 | 1502 | 2.0 | 63.8 | 0.05 |
| | | | | | EUROPE | | | | | | | |
| All | 7747 | 2.0 | 23.8 | 0.03 | 7068 | 2.1 | 41.0 | 0.04 | 7986 | 2.0 | 61.3 | 0.04 |
| >10 | 4155 | 2.0 | 23.8 | 0.03 | 4041 | 2.1 | 41.0 | 0.04 | 4474 | 2.0 | 61.3 | 0.04 |
| >mean | 1036 | 2.0 | 45.4 | 0.05 | 982 | 2.1 | 52.8 | 0.04 | 1040 | 2.0 | 63.7 | 0.05 |
| | | | | | NORTHERN AMERICA | | | | | | | |
| All | 9167 | 2.0 | 24.5 | 0.02 | 7562 | 2.0 | 28.9 | 0.02 | 6987 | 1.9 | 26.6 | 0.02 |
| >10 | 5222 | 2.0 | 24.5 | 0.03 | 4557 | 2.0 | 28.9 | 0.02 | 4115 | 1.9 | 26.6 | 0.02 |

|   |   |   |   |   |   |   |   |   |   |   |   |
|---|---|---|---|---|---|---|---|---|---|---|---|
| >mean | 1048 | 2.0 | 65.0 | 0.04 | 887 | 1.9 | 72.5 | 0.04 | 781 | 1.9 | 79.5 | 0.06 |
| SOUTHERN AMERICA | | | | | | | | | | | | |
| All | 2128 | 2.1 | 19.1 | 0.05 | 2513 | 2.0 | 16.9 | 0.06 | 3283 | 2.0 | 15.0 | 0.05 |
| >10 | 1207 | 2.1 | 19.1 | 0.05 | 1399 | 2.0 | 16.9 | 0.07 | 1711 | 2.0 | 15.0 | 0.05 |
| >mean | 339 | 2.1 | 39.7 | 0.07 | 402 | 2.1 | 40.1 | 0.08 | 493 | 2.0 | 39.6 | 0.07 |
| CHINA | | | | | | | | | | | | |
| All | 1389 | 2.1 | 24.1 | 0.13 | 1457 | 2.0 | 21.8 | 0.09 | 2437 | 2.0 | 33.9 | 0.07 |
| >10 | 727 | 2.1 | 24.1 | 0.14 | 865 | 2.0 | 21.8 | 0.10 | 1621 | 2.0 | 33.9 | 0.06 |
| >Mean | 252 | 2.2 | 34.1 | 0.21 | 239 | 2.1 | 49.5 | 0.11 | 340 | 2.1 | 71.3 | 0.09 |
| INDIA | | | | | | | | | | | | |
| All | 934 | 2.2 | 15.7 | 0.08 | 1094 | 2.1 | 15.5 | 0.08 | 1360 | 2.0 | 18.1 | 0.06 |
| >10 | 549 | 2.2 | 15.7 | 0.07 | 654 | 2.1 | 15.5 | 0.10 | 792 | 2.0 | 18.1 | 0.06 |
| >Mean | 184 | 2.2 | 30.4 | 0.16 | 208 | 2.1 | 34.6 | 0.25 | 234 | 2.1 | 40.5 | 0.26 |
| ISRAEL | | | | | | | | | | | | |
| All | 46 | 1.9 | 21.4 | 0.26 | 47 | 1.7 | 8.1 | 0.23 | 34 | 1.7 | 7.2 | 0.39 |
| >10 | 34 | 1.9 | 21.4 | 0.30 | 25 | 1.7 | 10.3 | 0.87 | 19 | 1.7 | 10.4 | 0.69 |
| NETHERLANDS | | | | | | | | | | | | |
| All | 142 | 2.2 | 29.2 | 0.32 | 99 | 2.5 | 32.8 | 0.59 | 78 | 2.3 | 44.9 | 0.48 |
| >10 | 72 | 2.2 | 29.2 | 0.36 | 65 | 2.5 | 32.8 | 0.61 | 48 | 2.3 | 44.9 | 0.49 |
| >Mean | 16 | 2.2 | 60.9 | 0.88 | 31 | 2.6 | 35.7 | N/A | 22 | 2.4 | 49.0 | 0.49 |
| ROMANIA | | | | | | | | | | | | |
| All | 73 | 3.7 | 46.4 | N/A | 93 | 3.7 | 74.3 | N/A | 118 | 2.2 | 25.5 | 0.30 |
| >10 | 36 | 3.7 | 46.4 | N/A | 59 | 3.7 | 74.3 | N/A | 82 | 2.2 | 25.5 | 0.29 |
| >Mean | 24 | 2.7 | 24.9 | N/A | 24 | 2.3 | 33.3 | N/A | 28 | 2.2 | 42.6 | N/A |
| RUSSIAN FEDERATION | | | | | | | | | | | | |
| All | 2942 | 2.0 | 35.0 | 0.10 | 1910 | 2.0 | 30.0 | 0.09 | 2229 | 2.1 | 34.8 | 0.12 |
| >10 | 1588 | 2.0 | 35.0 | 0.15 | 1190 | 2.0 | 30.0 | 0.09 | 1254 | 2.1 | 34.8 | 0.14 |
| >Mean | 523 | 2.0 | 45.2 | 0.34 | 360 | 2.1 | 49.0 | 0.23 | 402 | 2.1 | 45.6 | 0.26 |

Table 3: Top-ten countries according to number of cities for the three years considered (1992, 2001, and 2010)

| 1992 | | 2001 | | 2010 | |
|---|---|---|---|---|---|
| Country | # | country | # | country | # |
| UNITED STATES | 6576 | UNITED STATES | 5390 | UNITED STATES | 5160 |
| RUSSIAN FEDERATION | 2957 | RUSSIAN FEDERATION | 1910 | CHINA | 2437 |
| CANADA | 1689 | CHINA | 1457 | RUSSIAN FEDERATION | 2229 |
| CHINA | 1392 | BRAZIL | 1206 | BRAZIL | 1775 |
| BRAZIL | 1069 | CANADA | 1190 | INDIA | 1360 |
| INDIA | 934 | INDIA | 1094 | CANADA | 936 |
| GERMANY | 882 | FRANCE | 933 | FRANCE | 935 |
| FRANCE | 850 | GERMANY | 737 | GERMANY | 910 |
| ITALY | 829 | MEXICO | 719 | IRAN | 756 |
| MEXICO | 681 | ITALY | 712 | SPAIN | 731 |

We also found that city numbers among individual countries follow Zipf's law. That is, the number of natural cities in the first largest country is twice as many as that in the second largest country, three times as many as that in the third largest country, and so on. This regularity is well supported by the statistical tests, with Zipf's exponent between 0.9 and 1.0, and p values greater than 0.25. In addition, we observed that the number of natural cities in individual countries varies from one year to another; Table 3 shows the top-ten countries according to the number of natural cities during the three years; for example, the USA remained unchanged at the number one spot, although the number of cities has been decreasing, but China changed from number four to number two, with the number doubling within 20 years; in addition, Canada dropped significantly from number three to number six. How the rank changes reflect economic development remains an open issue for further study. For this purpose, we have made all the data publicly available to the scientific community.

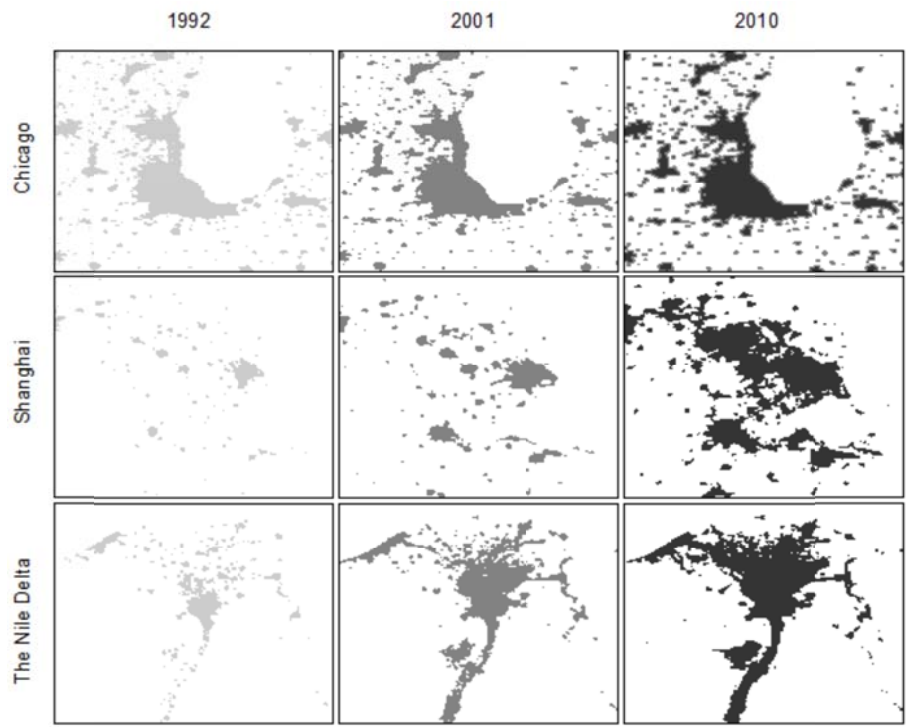

Figure 5: City growth near the regions of Shanghai, Chicago, and the Nile river valley and delta
(Note: There is little change in Chicago, but vast expansion in Shanghai and the Nile river regions.)

The 30,000 natural cities in the world provide a valuable data source for urban studies. Most of the natural cities correspond very well with real cities (for example Chicago), despite some exceptions (see Figure 5). For example, the cities near the Nile river valley and delta gradually merged into one natural city, where less than five percent of the land area within the country accommodates over 90 percent of the country's population. There are also some lit spots, such as burning wildfires, oil drilling, and burning natural gas, as well as volcano activities, but they were removed before our data was processed (Baugh et al. 2010). Within this study, for the first time, we have examined Zipf's law in a global setting involving all cities in the world. This is one of the defining advantages of working with big data (Mayer-Schonberger and Cukier 2013). The data, as an alternative to conventional statistical data, is particularly valuable for understanding the underlying mechanisms of urban structure and dynamics from the perspective of cities as a whole, rather than from that of individual cities.

**5. Further discussions on the study**
Zipf's law reveals an incredible regularity with respect to a set of cities that constitute an interconnected or interdependent whole, rather than for cities within a country, or an arbitrary set of cities. This study demonstrates the validity of Zipf's law at the global scale, at which the cities constitute a whole. The fact that Zipf's law does not hold for some continent or countries (or at some times) indicates that the cities within these geographic units do not constitute individual wholes. For example, all cities in Japan may not be a whole, given its global economic impacts; cities in some small countries are likely to be affected by neighboring or distant countries or regions, thus not exhibiting Zipf's law. For all cities of a country to constitute a whole, the country must usually be of a certain size. A country might not be a legitimate unit for Zipf's law. In the early days when Zipf's law was first discovered, there was little choice but to use cities in big countries for the investigation. Zipf (1949) adopted one hundred of the largest metropolitan districts in the USA in 1940 to show the regularity in a rank-size plot. At that time when data are scarce, researchers could not obtain data for all USA cities, let alone all cities in the world.

Natural cities are a product of the big data era, and can be extracted from remote sensing images, GPS and location-based social media data (Jiang and Miao 2014). Cities, as conventionally defined, are legitimate and convenient for administration and management, but might not be a good candidate for scientific inquiry due to their inexactness. For example, it was found that either incorporated places or metropolitan areas contain only 80 percent of the USA's population, while the remaining 20 percent is not counted in city-size distributions (Holmes and Lee 2010); theoretically speaking, all populations should be counted within individual cities or human settlements in general. Instead, accurately delineated natural cities are a better candidate for studying urban structure and dynamics (e.g., Schweitzer and Steinbrink 1997). This is because natural cities can be accurately delineated and measured, unlike conventional cities that are vaguely estimated. Thus, it is important to discuss some of the differences between conventional statistical data (which is obtained through censuses, sample surveys and statistical inferences) and big data (or digital transaction data, which is acquired and harvested from remote sensing, GPS and social media). The collection of statistical data such as that relating to populations is costly and labor intensive, so that it is only affordable to do so monthly, annually or once in a decade. This is very different from big data, or particularly social media data, which can be automatically acquired and can record events weekly, daily, and even down to minute and second scales. Conventional data and statistics have turned humanities into social sciences, but it is big data that is transforming social sciences into computational social science in the 21$^{st}$ century (Lazer et al. 2009). More importantly, conventional statistical data and big data represent two distinct ways of thinking: a centralized and top-down mindset within the former, and decentralized and bottom-up thinking within the latter.

The notion of natural cities implies a mindset shift from individual cities to cities as a whole – which is a different way of looking at cities. Cities from developing and developed countries play different roles, but they are essentially parts of the whole of cities. All the natural cities in the world constitute an interconnected or interdependent whole, just like human beings on the planet. Large cities or super-large cities emerged through mechanisms such as the "rich get richer" effect, and based on interaction and competition between cities. This view of cities as a whole (each of which consists of many cities) sets a clear difference from the view, in which cities are often studied individually in order to track their changes and growth. Under the view of cities as a whole (a network of cities), parts are self-similar to the whole. This is why Zipf's law holds in certain continents and countries, or at least at certain times. On the other hand, many countries, or many countries at some times, do not constitute valid parts, so Zipf's law in these countries is violated. To further illustrate the idea of parts and whole, we use the Koch curve, one of first fractals invented by the Swedish mathematician Helge von Koch (1870–1924). Obviously, the parts are exactly self-similar to the whole curve, but these parts must be correctly, rather than arbitrarily, identified (Figure 6) in order for this to be the case. In other words, the arbitrarily identified parts are not self-similar to the whole, even with the Koch curve. This is the same for cities. All cities in the world as a whole exhibit Zipf's law, but we need to identity correct parts in order to observe Zipf's law.

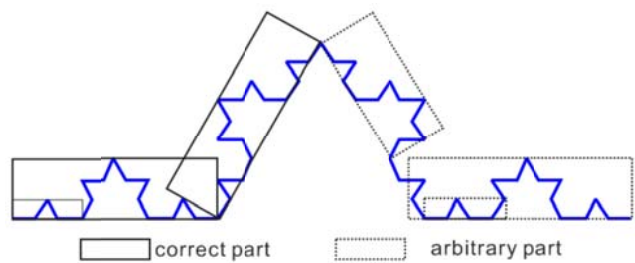

Figure 6: (Color online) Correctly versus arbitrarily identified parts of the Koch curve

The universality of Zipf's law, as argued and validated in this paper, adds some implications to building the new science of cities (Batty 2013). As argued earlier, top-down geographic data and units remain valid for many administrative decisions, such as elections and allocations of resources, but should not be the default choice for scientific investigation. First, while studying the underlying urban structure and dynamics (Batty 2005, Jiang and Yao 2010), we should not be constrained by the conventional census data, and should instead adopt geographic information acquired from remote sensing, or harvested from social media. There are major differences between the two kinds of data: the former is characterized as aggregated, sampled, and small in size, while the latter is individual, all-encompassing, and big in size. Second, we should not be constrained by census-bureau-imposed (or any top-down imposed) geographic units or boundaries, such as cities, counties, states, and countries, and instead should adopt naturally defined or delineated geographic units such as natural cities. Cities, and human activities in general, are naturally generated or self-organized through the interaction of people, which is not particularly constrained by country boundaries. This is especially true in the current circumstance when globalization is an irreversible trend. The failure to observe Zipf's law in some small countries occurs mainly because we are constrained by country boundaries. Also, while observing Zipf's law, we should not be constrained by particular time instants. This is in line with the dynamic (rather than static) view of looking the law or fractals (Jiang and Yin 2014). Third, cities are continuously and forever changing, and they grow and develop towards an idealized status, in which Zipf's law would appear. We should therefore be very cautious while arguing against the universality of Zipf's law.

## 6. Conclusion

This paper contended that Zipf's law is universal, with respect to the dispute on whether Zipf's law holds for a set of cities. Instead of using census-bureau-imposed definitions of cities, or human settlements, to be more precise, we considered natural cities extracted from night-time imagery in order to examine and verify Zipf's law in a global setting involving all natural cities in the world. It was found that Zipf's law holds remarkably well at the global level. However, it is violated in general at the country level, although it remains valid in certain places, and at certain times. Interestingly, among the six countries in which Zipf's law was not observed using real cities, we found that two of them exhibited Zipf's law very well at certain times. Based on these findings, we argued that Zipf's law applies to cities as a whole, rather than to an arbitrary set of cities, such as all cities in a country, and the continental and country levels are therefore not appropriate for observing Zipf's law in the big data era. Zipf's law is also reflected in the city numbers in individual countries. That is, the number of cities in the largest country is twice as many as that in the second largest country, three times as many as that in the third largest country, and so on.

We need new ways of thinking for studying cities, while facing big data harvested from remote sensing, GPS, and emerging social media. Cities are not just individual entities, but rather an interconnected or interdependent whole. To resolve the dispute on whether Zipf's law holds universally, we need correct data, scopes, and means of examining the power laws (such as maximum likelihood estimates). Conventional statistical data and top-down-imposed geographic units, as well as least squares estimates, often lead to questionable conclusions on Zipf's law. The 30,000 natural cities for each of the three years considered here could serve as benchmark data for further urban-related studies, thus contributing to the understanding of urban structure and dynamics. As seen in Table 3, there is a clear indicator that city expansions might be closely related to economic development. Our future work points to this direction.

**Acknowledgments**
We would like to thank Yun Jin for her assisting with some data processing, and Dr. Mark Newman for his insightful comments on PDF presented at Section 3. XXXX

**Appendix A: Power law detection for the 137 countries**

This appendix includes the power law detection results for 137 countries out of the 230 studied. For each country there are four parameters: the number of cities (#), the power law exponent (alpha, or alpha-1 as Zipf's exponent), the minimum above which the city sizes exhibit a power law, and an index indicating the goodness of fit (p). If p is set as N/A, then the corresponding power law is not trustworthy. The highlighted cells indicate that cities exhibit Zipf's law, while the empty cells imply no data available. For example, the number of cities in Afghanistan greater than 10 square kilometers or mean is fewer than six, which is too few to have reliable statistical tests.

|  | 1992 | | | | 2001 | | | | 2010 | | | |
|---|---|---|---|---|---|---|---|---|---|---|---|---|
|  | # | alpha | xmin | p | # | alpha | xmin | p | # | alpha | xmin | p |
| **AFGHANISTAN** | | | | | | | | | | | | |
| All | 9 | 1.7 | 5.2 | 0.84 | 7 | 1.7 | 3.7 | N/A | 20 | 2.9 | 38.6 | N/A |
| >10 | | | | | | | | | 13 | 2.9 | 38.6 | N/A |
| >Mean | | | | | | | | | 7 | 2.9 | 38.6 | N/A |
| **ALBANIA** | | | | | | | | | | | | |
| All | | | | | 9 | 2.5 | 11.7 | N/A | 14 | 1.6 | 2.6 | 0.45 |
| >10 | | | | | | | | | 7 | 2.4 | 20.4 | N/A |
| **ALGERIA** | | | | | | | | | | | | |
| All | 217 | 1.9 | 14.8 | 0.15 | 183 | 2.4 | 39.3 | 0.31 | 360 | 2.0 | 10.3 | 0.13 |
| >10 | 130 | 1.9 | 14.8 | 0.13 | 123 | 2.4 | 39.3 | 0.23 | 179 | 2.0 | 10.3 | 0.13 |
| >Mean | 37 | 2.0 | 47.7 | 0.38 | 40 | 2.4 | 39.3 | 0.36 | 67 | 2.1 | 30.8 | 0.39 |
| **ANGOLA** | | | | | | | | | | | | |
| All | 12 | 1.9 | 16.2 | N/A | 15 | 1.5 | 3.4 | 0.58 | 26 | 2.8 | 72.3 | 0.93 |
| >10 | 10 | 1.9 | 16.2 | 0.51 | 9 | 1.8 | 18.3 | N/A | 21 | 2.8 | 72.3 | N/A |
| >Mean | | | | | | | | | 7 | 2.8 | 72.3 | N/A |
| **ARGENTINA** | | | | | | | | | | | | |
| All | 368 | 2.0 | 13.3 | 0.16 | 526 | 2.1 | 19.9 | 0.10 | 647 | 2.1 | 14.4 | 0.09 |
| >10 | 191 | 2.0 | 13.3 | 0.17 | 273 | 2.1 | 19.9 | 0.08 | 324 | 2.1 | 14.4 | 0.10 |
| >Mean | 59 | 2.1 | 37.5 | 0.31 | 76 | 2.1 | 34.7 | 0.26 | 93 | 2.1 | 34.6 | 0.15 |
| **ARMENIA** | | | | | | | | | | | | |
| All | | | | | | | | | 8 | 1.8 | 5.2 | N/A |
| **AUSTRALIA** | | | | | | | | | | | | |
| All | 234 | 1.9 | 13.6 | 0.13 | 236 | 1.9 | 10.8 | 0.12 | 230 | 1.8 | 10.1 | 0.14 |
| >10 | 134 | 1.9 | 13.6 | 0.12 | 124 | 1.9 | 10.8 | 0.12 | 122 | 1.8 | 10.1 | 0.11 |
| >Mean | 30 | 2.0 | 59.3 | 0.23 | 28 | 1.9 | 58.7 | 0.56 | 28 | 1.9 | 67.3 | 0.18 |
| **AUSTRIA** | | | | | | | | | | | | |
| All | 117 | 2.1 | 10.5 | 0.21 | 117 | 2.0 | 11.2 | 0.20 | 132 | 2.1 | 21.5 | 0.34 |
| >10 | 58 | 2.1 | 10.5 | 0.25 | 55 | 2.0 | 11.2 | 0.21 | 74 | 2.1 | 21.5 | 0.32 |
| >Mean | 21 | 2.2 | 26.9 | N/A | 15 | 1.9 | 35.5 | 0.72 | 20 | 2.0 | 46.8 | 0.33 |
| **AZERBAIJAN** | | | | | | | | | | | | |
| All | 43 | 2.3 | 23.7 | 0.49 | 20 | 1.8 | 7.2 | 0.31 | 26 | 1.9 | 7.4 | 0.52 |
| >10 | 21 | 2.3 | 23.7 | 0.82 | 7 | 1.7 | 11.3 | N/A | 8 | 1.8 | 15.7 | N/A |
| >Mean | 7 | 2.4 | 40.2 | N/A | | | | | | | | |

|  |  |  |  |  |  |  |  |  |  |  |  |
|---|---|---|---|---|---|---|---|---|---|---|---|
|  |  |  |  | BANGLADESH |  |  |  |  |  |  |  |
| All | 54 | 2.1 | 7.8 | 0.35 | 39 | 2.2 | 14.9 | 0.47 | 33 | 2.1 | 21.7 | 0.45 |
| >10 | 26 | 2.2 | 18.9 | 0.54 | 22 | 2.2 | 14.9 | 0.37 | 20 | 2.1 | 21.7 | 0.64 |
| >Mean | 11 | 2.3 | 26.4 | 0.62 |  |  |  |  |  |  |  |  |
|  |  |  |  | BELARUS |  |  |  |  |  |  |  |
| All | 96 | 2.0 | 14.8 | 0.26 | 57 | 1.9 | 21.3 | 0.57 | 89 | 2.2 | 45.2 | 0.27 |
| >10 | 55 | 2.0 | 14.8 | 0.25 | 28 | 1.9 | 21.3 | 0.62 | 60 | 2.2 | 45.2 | 0.35 |
| >Mean | 21 | 2.3 | 39.9 | N/A | 13 | 2.1 | 39.8 | N/A | 20 | 2.2 | 51.2 | 0.41 |
|  |  |  |  | BELGIUM |  |  |  |  |  |  |  |
| All | 57 | 1.7 | 8.8 | 0.38 | 38 | 1.7 | 7.1 | 0.37 | 33 | 1.9 | 8.9 | N/A |
| >10 | 29 | 1.8 | 11.4 | 0.45 | 21 | 1.7 | 14.7 | 0.52 | 22 | 2.0 | 10.2 | 0.63 |
|  |  |  |  | BELIZE |  |  |  |  |  |  |  |
| All |  |  |  |  | 6 | 3.5 | 9.0 | N/A | 7 | 4.0 | 12.2 | N/A |
|  |  |  |  | BOLIVIA |  |  |  |  |  |  |  |
| All | 26 | 2.2 | 19.7 | 0.69 | 29 | 1.8 | 9.1 | 0.31 | 30 | 1.5 | 3.3 | 0.30 |
| >10 | 20 | 2.2 | 19.7 | 0.90 | 17 | 1.8 | 10.7 | 0.53 | 15 | 1.7 | 13.6 | N/A |
|  |  |  |  | BOSNIA AND HERZEGOVINA |  |  |  |  |  |  |  |
| All |  |  |  |  | 25 | 2.0 | 7.4 | N/A | 43 | 2.0 | 7.9 | 0.33 |
| >10 |  |  |  |  | 15 | 4.1 | 54.1 | N/A | 24 | 2.1 | 10.6 | 0.46 |
| >Mean |  |  |  |  | 7 | 2.2 | 26.8 | N/A | 10 | 2.2 | 26.9 | N/A |
|  |  |  |  | BOTSWANA |  |  |  |  |  |  |  |
| All | 9 | 2.8 | 14.1 | N/A | 13 | 2.5 | 12.5 | N/A | 14 | 2.3 | 9.5 | N/A |
| >10 |  |  |  |  | 9 | 2.5 | 12.5 | N/A | 10 | 2.3 | 10.2 | N/A |
|  |  |  |  | BRAZIL |  |  |  |  |  |  |  |
| All | 1068 | 2.1 | 18.0 | 0.08 | 1206 | 2.1 | 39.7 | 0.08 | 1775 | 2.0 | 16.5 | 0.07 |
| >10 | 594 | 2.1 | 18.0 | 0.07 | 649 | 2.1 | 39.7 | 0.09 | 904 | 2.0 | 16.5 | 0.05 |
| >Mean | 161 | 2.1 | 37.8 | 0.13 | 191 | 2.1 | 38.7 | 0.10 | 257 | 2.0 | 37.8 | 0.08 |
|  |  |  |  | BULGARIA |  |  |  |  |  |  |  |
| All | 62 | 2.8 | 23.0 | 0.77 | 39 | 2.5 | 15.9 | 0.57 | 63 | 2.1 | 14.2 | 0.33 |
| >10 | 34 | 2.8 | 23.0 | 0.86 | 28 | 2.5 | 15.9 | 0.64 | 36 | 2.1 | 14.2 | 0.37 |
| >Mean | 20 | 2.8 | 23.0 | 0.81 | 7 | 2.2 | 31.1 | N/A | 13 | 2.2 | 29.3 | 0.62 |
|  |  |  |  | CAMBODIA |  |  |  |  |  |  |  |
| All |  |  |  |  |  |  |  |  | 7 | 1.6 | 4.2 | N/A |
|  |  |  |  | CAMEROON |  |  |  |  |  |  |  |
| All | 12 | 1.5 | 0.9 | 0.33 | 7 | 2.1 | 22.1 | N/A | 7 | 1.4 | 2.5 | N/A |
|  |  |  |  | CANADA |  |  |  |  |  |  |  |
| All | 1684 | 2.0 | 18.9 | 0.05 | 1190 | 2.0 | 27.8 | 0.09 | 936 | 2.0 | 27.1 | 0.08 |
| >10 | 890 | 2.0 | 18.9 | 0.05 | 646 | 2.0 | 27.8 | 0.08 | 515 | 2.0 | 27.1 | 0.08 |
| >Mean | 209 | 2.0 | 53.3 | 0.11 | 156 | 2.0 | 55.9 | 0.13 | 117 | 2.0 | 56.8 | 0.12 |
|  |  |  |  | CHILE |  |  |  |  |  |  |  |
| All | 87 | 2.0 | 9.3 | 0.37 | 134 | 1.9 | 7.8 | 0.25 | 153 | 1.9 | 8.0 | 0.13 |
| >10 | 56 | 2.0 | 10.0 | 0.44 | 81 | 2.5 | 57.1 | 0.41 | 88 | 2.6 | 57.1 | 0.30 |
| >Mean | 19 | 2.4 | 40.2 | 0.83 | 27 | 2.2 | 39.1 | 0.66 | 29 | 2.2 | 39.1 | N/A |
|  |  |  |  | CHINA |  |  |  |  |  |  |  |
| All | 1389 | 2.1 | 24.1 | 0.13 | 1457 | 2.0 | 21.8 | 0.09 | 2437 | 2.0 | 33.9 | 0.07 |
| >10 | 727 | 2.1 | 24.1 | 0.14 | 865 | 2.0 | 21.8 | 0.10 | 1621 | 2.0 | 33.9 | 0.06 |
| >Mean | 252 | 2.2 | 34.1 | 0.21 | 239 | 2.1 | 49.5 | 0.11 | 340 | 2.1 | 71.3 | 0.09 |
|  |  |  |  | COLOMBIA |  |  |  |  |  |  |  |
| All | 182 | 1.9 | 11.0 | 0.12 | 181 | 1.9 | 19.8 | 0.21 | 195 | 1.9 | 12.7 | 0.12 |
| >10 | 96 | 1.9 | 11.0 | 0.13 | 104 | 1.9 | 19.8 | 0.20 | 98 | 1.9 | 12.7 | 0.14 |
| >Mean | 34 | 2.1 | 36.1 | 0.36 | 33 | 2.0 | 43.0 | 0.27 | 27 | 1.9 | 46.3 | 0.35 |
|  |  |  |  | CONGO_THE DEMOCRATIC REPUBLIC |  |  |  |  |  |  |  |
| All | 18 | 2.6 | 78.3 | 0.82 | 11 | 3.0 | 88.0 | N/A | 22 | 2.7 | 76.2 | 0.74 |
| >10 | 12 | 2.6 | 78.3 | 0.77 | 9 | 3.0 | 88.0 | N/A | 15 | 2.7 | 76.2 | 0.79 |
|  |  |  |  | COSTA RICA |  |  |  |  |  |  |  |
| All | 16 | 2.3 | 24.6 | N/A | 20 | 2.1 | 26.2 | 0.37 | 17 | 2.3 | 28.8 | N/A |
| >10 | 9 | 2.3 | 24.6 | N/A | 8 | 2.1 | 26.2 | N/A |  |  |  |  |
|  |  |  |  | COTE D'IVOIRE |  |  |  |  |  |  |  |
| All | 10 | 2.0 | 11.1 | N/A | 18 | 2.2 | 15.4 | N/A | 17 | 2.1 | 12.0 | 0.88 |
| >10 | 7 | 2.0 | 11.1 | N/A | 11 | 2.2 | 15.4 | 0.90 | 10 | 2.1 | 12.0 | 0.56 |
|  |  |  |  | CROATIA |  |  |  |  |  |  |  |
| All | 27 | 2.0 | 7.9 | 0.41 | 50 | 2.2 | 19.1 | N/A | 83 | 2.4 | 42.3 | 0.40 |
| >10 | 13 | 2.0 | 11.9 | 0.65 | 28 | 2.2 | 19.1 | 0.73 | 47 | 2.4 | 42.3 | 0.46 |

| | | | | | | | | | | | |
|---|---|---|---|---|---|---|---|---|---|---|---|
| >Mean | | | | | 11 | 2.5 | 36.9 | N/A | 17 | 2.4 | 44.7 | 0.51 |

CUBA

| | | | | | | | | | | | |
|---|---|---|---|---|---|---|---|---|---|---|---|
| All | 19 | 2.3 | 16.2 | 0.75 | 27 | 2.2 | 14.5 | 0.65 | 39 | 2.8 | 33.2 | N/A |
| >10 | 9 | 2.3 | 16.2 | N/A | 18 | 2.2 | 14.5 | 0.56 | 21 | 2.8 | 33.2 | N/A |
| >Mean | | | | | | | | | 12 | 2.8 | 32.4 | N/A |

CYPRUS

| | | | | | | | | | | | |
|---|---|---|---|---|---|---|---|---|---|---|---|
| All | 13 | 1.5 | 3.5 | N/A | 11 | 7.1 | 183.2 | N/A | 16 | 4.5 | 170.6 | N/A |
| >10 | 9 | 6.6 | 131.2 | N/A | 9 | 7.1 | 183.2 | N/A | 8 | 4.5 | 170.6 | N/A |

CZECH REPUBLIC

| | | | | | | | | | | | |
|---|---|---|---|---|---|---|---|---|---|---|---|
| All | 196 | 2.4 | 15.6 | 0.20 | 193 | 2.2 | 22.8 | 0.23 | 199 | 2.1 | 23.2 | 0.14 |
| >10 | 109 | 2.4 | 15.6 | 0.32 | 116 | 2.2 | 22.8 | 0.18 | 128 | 2.1 | 23.2 | 0.18 |
| >Mean | 49 | 2.5 | 24.1 | 0.94 | 33 | 2.2 | 38.7 | 0.62 | 37 | 2.1 | 46.6 | 0.53 |

DENMARK

| | | | | | | | | | | | |
|---|---|---|---|---|---|---|---|---|---|---|---|
| All | 99 | 2.2 | 22.6 | 0.32 | 59 | 2.0 | 23.2 | 0.24 | 121 | 2.0 | 17.1 | 0.20 |
| >10 | 54 | 2.2 | 22.6 | 0.31 | 37 | 2.0 | 23.2 | 0.27 | 69 | 2.0 | 17.1 | 0.23 |
| >Mean | 14 | 2.1 | 40.7 | 0.63 | 9 | 2.1 | 55.1 | 0.59 | 16 | 2.1 | 62.5 | 0.52 |

DOMINICAN REPUBLIC

| | | | | | | | | | | | |
|---|---|---|---|---|---|---|---|---|---|---|---|
| All | 22 | 2.2 | 16.3 | 0.81 | 32 | 2.2 | 23.6 | 0.54 | 24 | 2.0 | 15.4 | 0.55 |
| >10 | 17 | 2.2 | 16.3 | 0.56 | 20 | 2.2 | 23.6 | 0.78 | 14 | 2.0 | 15.4 | 0.77 |

ECUADOR

| | | | | | | | | | | | |
|---|---|---|---|---|---|---|---|---|---|---|---|
| All | 51 | 2.4 | 50.8 | 0.71 | 64 | 2.2 | 23.2 | 0.58 | 83 | 2.4 | 67.0 | 0.56 |
| >10 | 37 | 2.4 | 50.8 | 0.71 | 39 | 2.2 | 23.2 | 0.86 | 46 | 2.4 | 67.0 | 0.68 |
| >Mean | 14 | 2.4 | 50.8 | N/A | 13 | 2.3 | 48.9 | N/A | 19 | 2.2 | 51.6 | 0.83 |

EGYPT

| | | | | | | | | | | | |
|---|---|---|---|---|---|---|---|---|---|---|---|
| All | 196 | 1.9 | 16.9 | 0.22 | 162 | 1.8 | 16.8 | 0.14 | 140 | 1.8 | 11.0 | 0.15 |
| >10 | 134 | 1.9 | 16.9 | 0.17 | 100 | 1.8 | 16.8 | 0.16 | 77 | 1.8 | 11.0 | 0.19 |
| >Mean | 30 | 2.0 | 74.3 | 0.28 | 12 | 1.9 | 178.7 | 0.43 | | | | |

EL SALVADOR

| | | | | | | | | | | | |
|---|---|---|---|---|---|---|---|---|---|---|---|
| All | 14 | 1.7 | 2.5 | 0.35 | 19 | 2.0 | 13.4 | 0.89 | 16 | 1.9 | 6.7 | 0.53 |
| >10 | | | | | 12 | 2.0 | 13.4 | 0.65 | 11 | 2.0 | 10.0 | 0.47 |

ESTONIA

| | | | | | | | | | | | |
|---|---|---|---|---|---|---|---|---|---|---|---|
| All | 29 | 1.9 | 9.6 | 0.63 | 28 | 2.0 | 8.7 | 0.38 | 50 | 2.2 | 19.5 | 0.45 |
| >10 | 15 | 1.9 | 10.1 | 0.56 | 20 | 2.0 | 10.1 | N/A | 29 | 2.2 | 19.5 | 0.55 |
| >Mean | 7 | 2.2 | 28.5 | N/A | | | | | 9 | 2.1 | 45.1 | 0.59 |

ETHIOPIA

| | | | | | | | | | | | |
|---|---|---|---|---|---|---|---|---|---|---|---|
| All | | | | | 7 | 1.6 | 1.7 | N/A | 9 | 2.0 | 10.3 | N/A |

FINLAND

| | | | | | | | | | | | |
|---|---|---|---|---|---|---|---|---|---|---|---|
| All | 344 | 2.0 | 20.9 | 0.10 | 376 | 2.0 | 23.7 | 0.11 | 342 | 2.0 | 26.4 | 0.13 |
| >10 | 196 | 2.0 | 20.9 | 0.09 | 238 | 2.0 | 23.7 | 0.14 | 210 | 2.0 | 26.4 | 0.12 |
| >Mean | 60 | 2.0 | 36.2 | 0.44 | 63 | 2.0 | 47.7 | 0.47 | 49 | 1.9 | 56.5 | 0.55 |

FRANCE

| | | | | | | | | | | | |
|---|---|---|---|---|---|---|---|---|---|---|---|
| All | 850 | 1.9 | 12.2 | 0.13 | 933 | 2.1 | 59.5 | 0.14 | 935 | 2.1 | 69.2 | 0.11 |
| >10 | 447 | 1.9 | 12.2 | 0.11 | 506 | 2.1 | 59.5 | 0.13 | 502 | 2.1 | 69.2 | 0.15 |
| >Mean | 140 | 2.1 | 47.5 | 0.13 | 153 | 2.0 | 46.4 | 0.16 | 140 | 2.0 | 55.5 | 0.11 |

GABON

| | | | | | | | | | | | |
|---|---|---|---|---|---|---|---|---|---|---|---|
| All | 12 | 2.0 | 38.7 | 0.96 | 12 | 1.8 | 14.6 | 0.39 | 19 | 2.1 | 18.1 | 0.53 |
| >10 | 10 | 2.0 | 38.7 | N/A | 11 | 1.8 | 14.6 | 0.40 | 15 | 2.1 | 18.1 | 0.46 |

GEORGIA

| | | | | | | | | | | | |
|---|---|---|---|---|---|---|---|---|---|---|---|
| All | 18 | 2.1 | 9.7 | N/A | | | | | 12 | 2.4 | 35.2 | 0.90 |
| >10 | 9 | 2.1 | 10.9 | N/A | | | | | 9 | 2.4 | 35.2 | N/A |

GERMANY

| | | | | | | | | | | | |
|---|---|---|---|---|---|---|---|---|---|---|---|
| All | 880 | 2.0 | 13.3 | 0.07 | 737 | 2.0 | 50.3 | 0.12 | 910 | 1.9 | 19.6 | 0.15 |
| >10 | 451 | 2.0 | 13.3 | 0.07 | 437 | 2.0 | 50.3 | 0.11 | 519 | 1.9 | 19.6 | 0.14 |
| >Mean | 104 | 2.0 | 43.9 | 0.15 | 95 | 2.0 | 50.6 | 0.16 | 135 | 2.2 | 62.4 | 0.29 |

GHANA

| | | | | | | | | | | | |
|---|---|---|---|---|---|---|---|---|---|---|---|
| All | 13 | 3.3 | 98.5 | 0.72 | 14 | 1.8 | 15.4 | 0.36 | 15 | 1.9 | 20.5 | 0.52 |
| >10 | 9 | 3.3 | 98.5 | N/A | 9 | 1.8 | 15.4 | 0.53 | 11 | 1.9 | 20.5 | 0.41 |

GREECE

| | | | | | | | | | | | |
|---|---|---|---|---|---|---|---|---|---|---|---|
| All | 83 | 2.3 | 28.1 | 0.72 | 107 | 2.2 | 31.3 | 0.34 | 131 | 2.6 | 97.9 | 0.42 |
| >10 | 49 | 2.3 | 28.1 | 0.58 | 71 | 2.2 | 31.3 | 0.33 | 77 | 2.6 | 97.9 | 0.37 |
| >Mean | 14 | 2.3 | 43.1 | N/A | 20 | 2.3 | 55.8 | 0.82 | 26 | 2.2 | 54.5 | 0.48 |

GUATEMALA

| | | | | | | | | | | | |
|---|---|---|---|---|---|---|---|---|---|---|---|
| All | 18 | 2.2 | 11.7 | N/A | 26 | 1.9 | 6.7 | 0.65 | 24 | 1.9 | 5.0 | 0.44 |

|  |  |  |  |  |  |  |  |  |  |  |  |
|---|---|---|---|---|---|---|---|---|---|---|---|
| >10 | 8 | 2.2 | 11.7 | N/A | 12 | 2.1 | 18.3 | N/A | 11 | 2.1 | 15.0 | N/A |

HONDURAS

| | | | | | | | | | | | | |
|---|---|---|---|---|---|---|---|---|---|---|---|
| All | 8 | 1.7 | 6.7 | N/A | 18 | 1.8 | 11.7 | 0.50 | 19 | 1.8 | 6.6 | 0.33 |
| >10 | | | | | 11 | 1.8 | 11.7 | 0.58 | 11 | 2.0 | 21.7 | 0.70 |

HUNGARY

| | | | | | | | | | | | | |
|---|---|---|---|---|---|---|---|---|---|---|---|
| All | 87 | 2.7 | 32.7 | 0.38 | 104 | 2.1 | 13.7 | 0.47 | 104 | 2.2 | 16.1 | 0.24 |
| >10 | 53 | 2.7 | 32.7 | 0.64 | 60 | 2.1 | 13.7 | 0.52 | 60 | 2.2 | 16.1 | 0.40 |
| >Mean | 21 | 2.7 | 27.7 | N/A | 24 | 2.5 | 34.2 | N/A | 19 | 2.3 | 39.2 | N/A |

ICELAND

| | | | | | | | | | | | | |
|---|---|---|---|---|---|---|---|---|---|---|---|
| All | 41 | 2.1 | 8.6 | 0.53 | 17 | 1.8 | 5.7 | 0.38 | 26 | 2.2 | 11.2 | 0.46 |
| >10 | 16 | 2.2 | 14.4 | N/A | | | | | 16 | 2.2 | 11.2 | 0.70 |
| >Mean | 8 | 2.4 | 23.5 | N/A | | | | | | | | |

INDIA

| | | | | | | | | | | | | |
|---|---|---|---|---|---|---|---|---|---|---|---|
| All | 934 | 2.2 | 15.7 | 0.08 | 1094 | 2.1 | 15.5 | 0.08 | 1360 | 2.0 | 18.1 | 0.06 |
| >10 | 549 | 2.2 | 15.7 | 0.07 | 654 | 2.1 | 15.5 | 0.10 | 792 | 2.0 | 18.1 | 0.06 |
| >Mean | 184 | 2.2 | 30.4 | 0.16 | 208 | 2.1 | 34.6 | 0.25 | 234 | 2.1 | 40.5 | 0.26 |

INDONESIA

| | | | | | | | | | | | | |
|---|---|---|---|---|---|---|---|---|---|---|---|
| All | 183 | 1.9 | 11.9 | 0.40 | 187 | 1.8 | 12.8 | 0.25 | 207 | 1.9 | 12.9 | 0.08 |
| >10 | 119 | 1.9 | 11.9 | 0.38 | 108 | 1.8 | 12.8 | 0.33 | 130 | 1.9 | 12.9 | 0.10 |
| >Mean | 41 | 2.2 | 43.9 | 0.40 | 36 | 2.1 | 51.3 | 0.69 | 31 | 1.9 | 58.1 | 0.47 |

IRAN

| | | | | | | | | | | | | |
|---|---|---|---|---|---|---|---|---|---|---|---|
| All | 461 | 2.2 | 40.0 | 0.21 | 561 | 2.3 | 49.3 | 0.21 | 756 | 1.9 | 18.9 | 0.13 |
| >10 | 273 | 2.2 | 40.0 | 0.17 | 343 | 2.3 | 49.3 | 0.22 | 441 | 1.9 | 18.9 | 0.12 |
| >Mean | 63 | 2.2 | 54.1 | 0.22 | 108 | 2.2 | 44.1 | 0.18 | 134 | 2.1 | 51.9 | 0.15 |

IRAQ

| | | | | | | | | | | | | |
|---|---|---|---|---|---|---|---|---|---|---|---|
| All | 131 | 1.8 | 9.5 | 0.17 | 102 | 2.0 | 29.9 | 0.28 | 175 | 1.8 | 11.5 | 0.20 |
| >10 | 79 | 1.8 | 10.1 | 0.18 | 58 | 2.0 | 29.9 | 0.32 | 115 | 1.8 | 11.5 | 0.25 |
| >Mean | 24 | 2.1 | 57.3 | 0.46 | 16 | 2.0 | 74.7 | N/A | 28 | 2.2 | 75.9 | 0.49 |

IRELAND

| | | | | | | | | | | | | |
|---|---|---|---|---|---|---|---|---|---|---|---|
| All | 66 | 2.5 | 17.5 | 0.66 | 72 | 2.2 | 30.1 | 0.65 | 100 | 2.0 | 7.7 | 0.27 |
| >10 | 36 | 2.5 | 17.5 | 0.66 | 38 | 2.2 | 30.1 | 0.92 | 51 | 2.2 | 42.0 | 0.51 |
| >Mean | 9 | 2.1 | 29.2 | 0.73 | 11 | 2.2 | 33.4 | N/A | 17 | 2.5 | 39.0 | N/A |

ISRAEL

| | | | | | | | | | | | | |
|---|---|---|---|---|---|---|---|---|---|---|---|
| All | 46 | 1.9 | 21.4 | 0.26 | 47 | 1.7 | 8.1 | 0.23 | 34 | 1.7 | 7.2 | 0.39 |
| >10 | 34 | 1.9 | 21.4 | 0.30 | 25 | 1.7 | 10.3 | 0.87 | 19 | 1.7 | 10.4 | 0.69 |

ITALY

| | | | | | | | | | | | | |
|---|---|---|---|---|---|---|---|---|---|---|---|
| All | 826 | 2.0 | 29.4 | 0.10 | 712 | 2.0 | 44.3 | 0.14 | 664 | 1.9 | 47.3 | 0.14 |
| >10 | 446 | 2.0 | 29.4 | 0.10 | 410 | 2.0 | 44.3 | 0.13 | 349 | 1.9 | 47.3 | 0.12 |
| >Mean | 105 | 2.0 | 56.7 | 0.14 | 83 | 2.0 | 76.6 | 0.21 | 65 | 1.9 | 102.6 | 0.24 |

JAMAICA

| | | | | | | | | | | | | |
|---|---|---|---|---|---|---|---|---|---|---|---|
| All | 15 | 2.2 | 27.0 | N/A | 14 | 2.2 | 31.8 | N/A | 9 | 2.0 | 16.3 | 0.88 |
| >10 | 12 | 2.2 | 27.0 | N/A | 10 | 2.2 | 31.8 | 0.79 | 8 | 2.0 | 16.3 | N/A |

JAPAN

| | | | | | | | | | | | | |
|---|---|---|---|---|---|---|---|---|---|---|---|
| All | 572 | 1.8 | 21.1 | 0.07 | 491 | 1.7 | 10.0 | 0.08 | 423 | 1.8 | 27.7 | 0.14 |
| >10 | 324 | 1.8 | 21.1 | 0.06 | 324 | 1.7 | 10.0 | 0.06 | 271 | 1.8 | 27.7 | 0.10 |
| >Mean | 59 | 1.9 | 127.3 | 0.32 | 57 | 1.9 | 135.0 | 0.33 | 49 | 1.9 | 151.0 | 0.46 |

JORDAN

| | | | | | | | | | | | | |
|---|---|---|---|---|---|---|---|---|---|---|---|
| All | 28 | 2.0 | 15.5 | 0.51 | 38 | 2.0 | 13.3 | 0.36 | 41 | 1.7 | 10.3 | 0.26 |
| >10 | 14 | 2.0 | 15.5 | 0.61 | 23 | 2.0 | 13.3 | 0.57 | 27 | 1.7 | 10.3 | 0.26 |
| >Mean | | | | | 7 | 2.2 | 51.7 | N/A | | | | |

KAZAKHSTAN

| | | | | | | | | | | | | |
|---|---|---|---|---|---|---|---|---|---|---|---|
| All | 284 | 1.9 | 11.0 | 0.20 | 96 | 2.2 | 49.4 | 0.61 | 175 | 1.6 | 4.6 | 0.41 |
| >10 | 138 | 1.9 | 11.0 | 0.30 | 66 | 2.2 | 49.4 | 0.52 | 87 | 3.1 | 142.9 | 0.53 |
| >Mean | 57 | 2.0 | 31.4 | 0.94 | 26 | 2.3 | 58.9 | N/A | 41 | 2.0 | 41.8 | 0.58 |

KENYA

| | | | | | | | | | | | | |
|---|---|---|---|---|---|---|---|---|---|---|---|
| All | 10 | 1.8 | 6.0 | N/A | | | | | 6 | 1.4 | 1.7 | 0.83 |
| >10 | 7 | 2.0 | 11.2 | N/A | | | | | | | | |

KOREA_DEMOCRATIC PEOPLE'S RE

| | | | | | | | | | | | | |
|---|---|---|---|---|---|---|---|---|---|---|---|
| All | 8 | 1.4 | 0.7 | N/A | 10 | 4.3 | 32.1 | N/A | 10 | 1.8 | 7.4 | 0.85 |
| >10 | | | | | | | | | 7 | 2.1 | 14.3 | N/A |

KOREA_REPUBLIC OF

| | | | | | | | | | | | | |
|---|---|---|---|---|---|---|---|---|---|---|---|
| All | 170 | 1.8 | 14.4 | 0.12 | 171 | 1.8 | 12.3 | 0.12 | 167 | 1.7 | 19.5 | 0.11 |
| >10 | 97 | 1.8 | 14.4 | 0.11 | 102 | 1.8 | 12.3 | 0.13 | 108 | 1.7 | 19.5 | 0.11 |
| >Mean | 20 | 1.9 | 82.3 | 0.80 | 16 | 1.8 | 132.7 | 0.59 | 18 | 1.9 | 161.1 | 0.74 |

KUWAIT

| | | | | | | | | | | | | |
|---|---|---|---|---|---|---|---|---|---|---|---|---|
| All | 10 | 1.7 | 29.8 | 0.78 | 14 | 1.6 | 16.4 | 0.43 | 14 | 1.6 | 14.3 | 0.40 |
| >10 | | | | | 10 | 1.6 | 16.4 | 0.56 | 10 | 1.6 | 14.3 | 0.58 |

KYRGYZSTAN

| | | | | | | | | | | | | |
|---|---|---|---|---|---|---|---|---|---|---|---|---|
| All | 32 | 2.4 | 13.6 | 0.71 | 20 | 2.1 | 7.2 | 0.47 | 25 | 2.2 | 12.6 | 0.79 |
| >10 | 19 | 2.4 | 13.6 | 0.89 | 8 | 2.2 | 13.0 | N/A | 15 | 2.2 | 12.6 | N/A |
| >Mean | 9 | 2.6 | 22.9 | 0.84 | | | | | | | | |

LAO PEOPLE'S DEMOCRATIC REPUBL

| | | | | | | | | | | | | |
|---|---|---|---|---|---|---|---|---|---|---|---|---|
| All | | | | | | | | | 9 | 2.5 | 41.9 | 0.89 |
| >10 | | | | | | | | | 8 | 2.5 | 41.9 | N/A |

LATVIA

| | | | | | | | | | | | | |
|---|---|---|---|---|---|---|---|---|---|---|---|---|
| All | 23 | 1.6 | 1.9 | N/A | 24 | 2.8 | 29.9 | N/A | 39 | 2.5 | 33.0 | N/A |
| >10 | 10 | 2.3 | 16.6 | N/A | 12 | 2.8 | 29.9 | N/A | 24 | 2.5 | 33.0 | N/A |
| >Mean | | | | | 7 | 2.8 | 29.9 | N/A | | | | |

LEBANON

| | | | | | | | | | | | | |
|---|---|---|---|---|---|---|---|---|---|---|---|---|
| All | 13 | 2.5 | 46.8 | N/A | 22 | 1.6 | 5.0 | 0.22 | 16 | 1.7 | 22.2 | 0.45 |
| >10 | 7 | 2.5 | 46.8 | N/A | 14 | 1.7 | 22.2 | 0.36 | 10 | 1.7 | 22.2 | N/A |

LIBYA_ARAB JAMAHIRIY_

| | | | | | | | | | | | | |
|---|---|---|---|---|---|---|---|---|---|---|---|---|
| All | 156 | 1.9 | 14.5 | 0.10 | 157 | 2.0 | 10.5 | 0.24 | 197 | 2.1 | 30.1 | 0.24 |
| >10 | 103 | 1.9 | 14.5 | 0.11 | 107 | 2.0 | 10.5 | 0.28 | 114 | 2.1 | 30.1 | 0.21 |
| >Mean | 25 | 1.9 | 57.9 | 1.00 | 30 | 2.2 | 42.0 | 0.46 | 31 | 2.1 | 51.5 | 0.44 |

LITHUANIA

| | | | | | | | | | | | | |
|---|---|---|---|---|---|---|---|---|---|---|---|---|
| All | 36 | 2.3 | 18.5 | 0.50 | 26 | 2.2 | 16.1 | 0.40 | 53 | 2.0 | 13.0 | 0.36 |
| >10 | 18 | 2.3 | 18.5 | 0.50 | 13 | 2.2 | 16.1 | 0.43 | 37 | 2.0 | 13.0 | 0.36 |
| >Mean | 9 | 2.3 | 25.8 | 0.81 | | | | | 11 | 2.2 | 45.2 | N/A |

LUXEMBOURG

| | | | | | | | | | | | | |
|---|---|---|---|---|---|---|---|---|---|---|---|---|
| All | 9 | 1.8 | 13.4 | 0.69 | 10 | 1.7 | 5.5 | 0.37 | 6 | 1.4 | 2.2 | 0.29 |

MACEDONIA_THE FORMER YUGOSLAV

| | | | | | | | | | | | | |
|---|---|---|---|---|---|---|---|---|---|---|---|---|
| All | 13 | 2.8 | 12.2 | N/A | 20 | 2.8 | 18.6 | N/A | 18 | 2.6 | 17.2 | N/A |
| >10 | 8 | 2.8 | 12.2 | N/A | 13 | 2.8 | 18.6 | N/A | 11 | 2.6 | 17.2 | N/A |

MADAGASCAR

| | | | | | | | | | | | | |
|---|---|---|---|---|---|---|---|---|---|---|---|---|
| All | | | | | | | | | 6 | 2.1 | 4.9 | N/A |

MALAYSIA

| | | | | | | | | | | | | |
|---|---|---|---|---|---|---|---|---|---|---|---|---|
| All | 61 | 2.5 | 66.9 | 0.89 | 104 | 2.1 | 66.2 | 0.34 | 133 | 1.7 | 9.4 | 0.17 |
| >10 | 38 | 2.5 | 66.9 | N/A | 63 | 2.1 | 66.2 | 0.44 | 92 | 1.7 | 10.3 | 0.20 |
| >Mean | 16 | 2.4 | 56.5 | 0.89 | 17 | 2.1 | 96.3 | 0.46 | 19 | 2.0 | 122.1 | 0.46 |

MALI

| | | | | | | | | | | | | |
|---|---|---|---|---|---|---|---|---|---|---|---|---|
| All | | | | | | | | | 9 | 1.9 | 4.1 | 0.90 |

MEXICO

| | | | | | | | | | | | | |
|---|---|---|---|---|---|---|---|---|---|---|---|---|
| All | 679 | 1.9 | 10.4 | 0.06 | 719 | 1.9 | 10.4 | 0.06 | 652 | 1.8 | 10.4 | 0.06 |
| >10 | 356 | 1.9 | 10.4 | 0.08 | 393 | 1.9 | 10.4 | 0.06 | 351 | 1.8 | 10.4 | 0.05 |
| >Mean | 109 | 2.0 | 42.1 | 0.27 | 104 | 1.9 | 46.9 | 0.36 | 91 | 1.8 | 59.5 | 0.37 |

MOLDOVA_REPUBLIC OF

| | | | | | | | | | | | | |
|---|---|---|---|---|---|---|---|---|---|---|---|---|
| All | 41 | 2.3 | 14.7 | 0.54 | | | | | 6 | 1.7 | 7.6 | N/A |
| >10 | 21 | 2.3 | 14.7 | 0.46 | | | | | | | | |
| >Mean | 8 | 2.2 | 23.0 | 0.88 | | | | | | | | |

MONGOLIA

| | | | | | | | | | | | | |
|---|---|---|---|---|---|---|---|---|---|---|---|---|
| All | 8 | 1.4 | 0.6 | 0.44 | | | | | 10 | 2.0 | 9.4 | N/A |

MOROCCO

| | | | | | | | | | | | | |
|---|---|---|---|---|---|---|---|---|---|---|---|---|
| All | 81 | 2.2 | 11.2 | 0.21 | 90 | 2.0 | 11.5 | 0.24 | 107 | 1.9 | 9.3 | 0.16 |
| >10 | 56 | 2.2 | 11.2 | 0.21 | 57 | 2.0 | 11.5 | 0.17 | 70 | 1.9 | 10.0 | 0.17 |
| >Mean | 15 | 2.0 | 32.5 | 0.70 | 18 | 2.1 | 33.9 | N/A | 14 | 1.7 | 42.2 | N/A |

MOZAMBIQUE

| | | | | | | | | | | | | |
|---|---|---|---|---|---|---|---|---|---|---|---|---|
| All | 8 | 3.3 | 30.2 | N/A | 8 | 1.7 | 3.3 | 0.47 | 13 | 2.0 | 18.8 | 0.78 |
| >10 | | | | | | | | | 9 | 2.0 | 18.8 | N/A |

MYANMAR

| | | | | | | | | | | | | |
|---|---|---|---|---|---|---|---|---|---|---|---|---|
| All | 10 | 1.8 | 4.0 | 0.35 | 13 | 1.7 | 5.7 | 0.59 | 16 | 1.6 | 4.0 | 0.31 |
| >10 | | | | | | | | | 8 | 1.7 | 10.2 | N/A |

NAMIBIA

| | | | | | | | | | | | | |
|---|---|---|---|---|---|---|---|---|---|---|---|---|
| All | 15 | 2.8 | 14.5 | N/A | 17 | 2.1 | 10.0 | 0.75 | 14 | 1.9 | 6.1 | N/A |
| >10 | 9 | 2.8 | 14.5 | N/A | | | | | 7 | 2.0 | 10.3 | N/A |

NEPAL

|       |     |     |      |      |     |     |       |      |     |     |       |      |
|-------|-----|-----|------|------|-----|-----|-------|------|-----|-----|-------|------|
| All   | 6   | 2.5 | 13.9 | N/A  |     |     |       |      |     |     |       |      |
|       |     |     |      |      |     | NETHERLANDS |   |      |     |     |       |      |
| All   | 142 | 2.2 | 29.2 | 0.32 | 99  | 2.5 | 32.8  | 0.59 | 78  | 2.3 | 44.9  | 0.48 |
| >10   | 72  | 2.2 | 29.2 | 0.36 | 65  | 2.5 | 32.8  | 0.61 | 48  | 2.3 | 44.9  | 0.49 |
| >Mean | 16  | 2.2 | 60.9 | 0.88 | 31  | 2.6 | 35.7  | N/A  | 22  | 2.4 | 49.0  | 0.49 |
|       |     |     |      |      |     | NEW ZEALAND |   |      |     |     |       |      |
| All   | 42  | 2.1 | 25.0 | 0.37 | 36  | 1.7 | 6.2   | 0.39 | 39  | 1.6 | 5.4   | 0.61 |
| >10   | 27  | 2.1 | 25.0 | N/A  | 23  | 1.8 | 13.3  | 0.54 | 24  | 2.1 | 37.0  | 0.45 |
| >Mean | 7   | 2.2 | 64.9 | N/A  |     |     |       |      | 8   | 2.1 | 60.4  | N/A  |
|       |     |     |      |      |     | NICARAGUA |     |      |     |     |       |      |
| All   | 10  | 2.3 | 10.9 | N/A  | 15  | 2.3 | 15.1  | 0.84 | 14  | 2.9 | 18.5  | N/A  |
| >10   |     |     |      |      | 7   | 2.3 | 15.1  | N/A  | 8   | 2.9 | 18.5  | N/A  |
|       |     |     |      |      |     | NIGER |       |      |     |     |       |      |
| All   | 7   | 2.5 | 7.3  | N/A  |     |     |       |      | 7   | 2.1 | 5.8   | N/A  |
|       |     |     |      |      |     | NIGERIA |     |      |     |     |       |      |
| All   | 87  | 1.8 | 67.8 | 0.25 | 82  | 1.7 | 23.1  | 0.19 | 107 | 2.0 | 45.4  | 0.44 |
| >10   | 59  | 1.8 | 67.8 | 0.54 | 64  | 1.7 | 23.1  | 0.25 | 80  | 2.0 | 45.4  | 0.27 |
| >Mean | 9   | 1.8 | 253.9| N/A  | 13  | 2.0 | 221.3 | N/A  | 21  | 2.1 | 101.2 | 0.58 |
|       |     |     |      |      |     | NORWAY |      |      |     |     |       |      |
| All   | 234 | 2.0 | 17.1 | 0.12 | 256 | 2.0 | 17.3  | 0.13 | 265 | 1.9 | 12.7  | 0.11 |
| >10   | 121 | 2.0 | 17.1 | 0.14 | 133 | 2.0 | 17.3  | 0.16 | 139 | 1.9 | 12.7  | 0.11 |
| >Mean | 39  | 2.0 | 39.6 | 0.41 | 39  | 2.1 | 46.8  | 0.39 | 30  | 1.9 | 59.0  | 0.30 |
|       |     |     |      |      |     | OMAN |       |      |     |     |       |      |
| All   | 76  | 2.1 | 14.7 | 0.23 | 90  | 2.0 | 18.2  | 0.21 | 109 | 1.7 | 6.3   | 0.22 |
| >10   | 54  | 2.1 | 14.7 | 0.25 | 65  | 2.0 | 18.2  | 0.20 | 67  | 1.8 | 10.3  | 0.45 |
| >Mean | 10  | 2.0 | 66.1 | 0.64 |     |     |       |      |     |     |       |      |
|       |     |     |      |      |     | PAKISTAN |    |      |     |     |       |      |
| All   | 305 | 2.3 | 17.5 | 0.17 | 300 | 2.2 | 20.2  | 0.16 | 245 | 2.0 | 11.4  | 0.13 |
| >10   | 180 | 2.3 | 17.5 | 0.15 | 168 | 2.2 | 20.2  | 0.15 | 125 | 2.0 | 11.4  | 0.12 |
| >Mean | 62  | 2.3 | 29.4 | 0.27 | 58  | 2.2 | 31.2  | 0.20 | 33  | 2.0 | 37.7  | 0.27 |
|       |     |     |      |      |     | PALESTINE |   |      |     |     |       |      |
| All   | 25  | 2.4 | 30.5 | N/A  | 21  | 1.6 | 5.9   | 0.34 | 20  | 1.6 | 2.9   | 0.48 |
| >10   | 16  | 2.4 | 30.5 | 0.84 | 12  | 1.6 | 10.2  | 0.40 | 7   | 1.6 | 19.6  | N/A  |
|       |     |     |      |      |     | PANAMA |      |      |     |     |       |      |
| All   | 13  | 2.1 | 17.9 | N/A  | 13  | 1.8 | 10.2  | N/A  | 11  | 1.9 | 21.3  | 0.45 |
| >10   | 8   | 2.1 | 17.9 | N/A  | 11  | 1.8 | 10.2  | 0.89 | 8   | 1.9 | 21.3  | 0.47 |
|       |     |     |      |      |     | PAPUA NEW GUINEA | | |   |     |     |       |      |
| All   | 11  | 2.0 | 6.0  | N/A  | 7   | 1.5 | 1.7   | N/A  | 6   | 2.3 | 9.4   | N/A  |
|       |     |     |      |      |     | PARAGUAY |    |      |     |     |       |      |
| All   | 14  | 1.6 | 7.1  | 0.38 | 25  | 1.7 | 10.0  | 0.27 | 39  | 1.8 | 8.0   | 0.41 |
| >10   | 9   | 1.7 | 10.9 | 0.49 | 13  | 1.7 | 10.9  | 0.45 | 20  | 1.8 | 10.9  | 0.59 |
|       |     |     |      |      |     | PERU |       |      |     |     |       |      |
| All   | 48  | 2.1 | 10.9 | 0.46 | 73  | 1.9 | 8.4   | 0.77 | 88  | 2.5 | 58.1  | 0.40 |
| >10   | 35  | 2.1 | 10.9 | 0.47 | 43  | 2.0 | 16.9  | 0.80 | 50  | 2.5 | 58.1  | 0.50 |
| >Mean | 9   | 2.4 | 46.8 | N/A  | 13  | 2.2 | 43.5  | N/A  | 17  | 2.2 | 47.6  | 0.80 |
|       |     |     |      |      |     | PHILIPPINES | |      |     |     |       |      |
| All   | 30  | 2.2 | 35.7 | 0.74 | 39  | 1.7 | 6.7   | 0.26 | 42  | 1.7 | 8.3   | 0.48 |
| >10   | 16  | 2.2 | 35.7 | N/A  | 25  | 1.7 | 11.0  | 0.27 | 29  | 1.8 | 10.0  | 0.55 |
|       |     |     |      |      |     | POLAND |      |      |     |     |       |      |
| All   | 338 | 2.1 | 11.9 | 0.15 | 369 | 2.2 | 27.6  | 0.15 | 537 | 2.1 | 67.2  | 0.15 |
| >10   | 186 | 2.1 | 11.9 | 0.15 | 232 | 2.2 | 27.6  | 0.15 | 356 | 2.1 | 67.2  | 0.15 |
| >Mean | 57  | 2.3 | 35.9 | 0.34 | 65  | 2.2 | 47.5  | 0.23 | 87  | 2.2 | 71.0  | 0.17 |
|       |     |     |      |      |     | PORTUGAL |   |      |     |     |       |      |
| All   | 108 | 2.1 | 12.9 | 0.22 | 168 | 2.0 | 16.0  | 0.17 | 169 | 1.9 | 20.6  | 0.22 |
| >10   | 59  | 2.1 | 12.9 | 0.20 | 90  | 2.0 | 16.0  | 0.14 | 83  | 1.9 | 20.6  | 0.23 |
| >Mean | 12  | 1.9 | 51.6 | 0.51 | 20  | 2.1 | 60.4  | 0.45 | 17  | 2.0 | 75.6  | 0.75 |
|       |     |     |      |      |     | PUERTO RICO | |      |     |     |       |      |
| All   | 20  | 1.5 | 8.2  | 0.27 | 13  | 1.5 | 9.8   | 0.27 | 12  | 1.4 | 4.9   | N/A  |
| >10   | 14  | 1.6 | 11.4 | 0.27 | 9   | 1.4 | 11.4  | 0.41 | 10  | 1.5 | 11.4  | N/A  |
|       |     |     |      |      |     | QATAR |      |      |     |     |       |      |
| All   | 16  | 1.6 | 7.8  | 0.26 | 15  | 2.5 | 164.6 | 0.88 | 8   | 1.7 | 7.0   | N/A  |
| >10   | 9   | 1.5 | 10.9 | 0.40 | 9   | 2.5 | 164.6 | N/A  |     |     |       |      |
|       |     |     |      |      |     | REUNION |    |      |     |     |       |      |

|       | | | | | | | | | | | |
|-------|---|---|---|---|---|---|---|---|---|---|---|
| All   | 8 | 1.9 | 8.8 | 0.90 | 6 | 6.0 | 83.6 | N/A | 6 | 2.5 | 66.7 | N/A |

ROMANIA

|       | | | | | | | | | | | |
|-------|---|---|---|---|---|---|---|---|---|---|---|
| All   | 73 | 3.7 | 46.4 | N/A | 93 | 3.7 | 74.3 | N/A | 118 | 2.2 | 25.5 | 0.30 |
| >10   | 36 | 3.7 | 46.4 | N/A | 59 | 3.7 | 74.3 | N/A | 82 | 2.2 | 25.5 | 0.29 |
| >Mean | 24 | 2.7 | 24.9 | N/A | 24 | 2.3 | 33.3 | N/A | 28 | 2.2 | 42.6 | N/A |

RUSSIAN FEDERATION

|       | | | | | | | | | | | |
|-------|---|---|---|---|---|---|---|---|---|---|---|
| All   | 2942 | 2.0 | 35.0 | 0.10 | 1910 | 2.0 | 30.0 | 0.09 | 2229 | 2.1 | 34.8 | 0.12 |
| >10   | 1588 | 2.0 | 35.0 | 0.15 | 1190 | 2.0 | 30.0 | 0.09 | 1254 | 2.1 | 34.8 | 0.14 |
| >Mean | 523 | 2.0 | 45.2 | 0.34 | 360 | 2.1 | 49.0 | 0.23 | 402 | 2.1 | 45.6 | 0.26 |

SAUDI ARABIA

|       | | | | | | | | | | | |
|-------|---|---|---|---|---|---|---|---|---|---|---|
| All   | 378 | 1.9 | 18.5 | 0.10 | 373 | 2.0 | 29.3 | 0.09 | 568 | 2.0 | 50.3 | 0.18 |
| >10   | 234 | 1.9 | 18.5 | 0.09 | 253 | 2.0 | 29.3 | 0.10 | 341 | 2.0 | 50.3 | 0.14 |
| >Mean | 49 | 1.9 | 72.0 | 0.31 | 52 | 1.9 | 69.2 | 0.31 | 93 | 2.1 | 70.2 | 0.14 |

SENEGAL

|       | | | | | | | | | | | |
|-------|---|---|---|---|---|---|---|---|---|---|---|
| All   | 7 | 2.0 | 5.8 | N/A | 8 | 2.1 | 13.4 | N/A | 8 | 1.7 | 6.6 | 0.45 |

SLOVAKIA

|       | | | | | | | | | | | |
|-------|---|---|---|---|---|---|---|---|---|---|---|
| All   | 116 | 2.2 | 16.0 | 0.42 | 79 | 3.5 | 56.9 | 0.55 | 69 | 2.3 | 26.6 | 0.42 |
| >10   | 66 | 2.2 | 16.0 | 0.39 | 55 | 3.5 | 56.9 | 0.95 | 47 | 2.3 | 26.6 | 0.38 |
| >Mean | 30 | 2.3 | 25.2 | 0.93 | 23 | 2.6 | 28.3 | N/A | 19 | 2.4 | 36.7 | N/A |

SLOVENIA

|       | | | | | | | | | | | |
|-------|---|---|---|---|---|---|---|---|---|---|---|
| All   | 23 | 2.0 | 10.2 | 0.51 | 25 | 1.9 | 17.9 | 0.31 | 35 | 1.7 | 4.1 | 0.52 |
| >10   | 13 | 2.0 | 10.2 | 0.45 | 15 | 1.9 | 17.9 | 0.40 | 17 | 2.1 | 21.7 | 0.59 |
| >Mean | | | | | | | | | | | | |

SOUTH AFRICA

|       | | | | | | | | | | | |
|-------|---|---|---|---|---|---|---|---|---|---|---|
| All   | 203 | 1.9 | 14.7 | 0.11 | 238 | 1.9 | 11.8 | 0.13 | 254 | 2.0 | 16.2 | 0.15 |
| >10   | 136 | 1.9 | 14.7 | 0.14 | 149 | 1.9 | 11.8 | 0.18 | 144 | 2.0 | 16.2 | 0.14 |
| >Mean | 24 | 1.9 | 79.2 | 0.47 | 30 | 2.0 | 65.8 | 0.90 | 34 | 2.1 | 62.8 | 0.32 |

SPAIN

|       | | | | | | | | | | | |
|-------|---|---|---|---|---|---|---|---|---|---|---|
| All   | 618 | 1.9 | 10.0 | 0.06 | 703 | 1.9 | 12.5 | 0.07 | 731 | 1.8 | 9.3 | 0.06 |
| >10   | 333 | 1.9 | 10.0 | 0.06 | 379 | 1.9 | 12.5 | 0.07 | 363 | 1.9 | 14.3 | 0.06 |
| >Mean | 85 | 1.9 | 48.3 | 0.17 | 86 | 1.9 | 52.0 | 0.29 | 77 | 1.8 | 60.1 | 0.20 |

SRI LANKA

|       | | | | | | | | | | | |
|-------|---|---|---|---|---|---|---|---|---|---|---|
| All   | 6 | 1.9 | 7.7 | N/A | 10 | 1.9 | 5.1 | N/A | 17 | 1.8 | 5.1 | N/A |

SUDAN

|       | | | | | | | | | | | |
|-------|---|---|---|---|---|---|---|---|---|---|---|
| All   | 30 | 2.2 | 12.5 | 0.98 | 27 | 2.2 | 15.1 | 0.64 | 47 | 2.3 | 31.8 | 0.59 |
| >10   | 15 | 2.2 | 12.5 | 0.85 | 12 | 2.2 | 15.1 | N/A | 26 | 2.3 | 31.8 | N/A |
| >Mean | | | | | | | | | 8 | 2.3 | 41.3 | N/A |

SWEDEN

|       | | | | | | | | | | | |
|-------|---|---|---|---|---|---|---|---|---|---|---|
| All   | 543 | 2.1 | 12.2 | 0.10 | 412 | 2.2 | 22.5 | 0.12 | 460 | 2.1 | 31.2 | 0.14 |
| >10   | 315 | 2.1 | 12.2 | 0.11 | 230 | 2.2 | 22.5 | 0.16 | 269 | 2.1 | 31.2 | 0.13 |
| >Mean | 114 | 2.3 | 30.9 | 0.18 | 87 | 2.2 | 34.2 | 0.32 | 82 | 2.1 | 51.7 | 0.33 |

SWITZERLAND

|       | | | | | | | | | | | |
|-------|---|---|---|---|---|---|---|---|---|---|---|
| All   | 117 | 1.7 | 4.7 | 0.25 | 104 | 2.1 | 36.5 | 0.33 | 70 | 1.7 | 8.8 | 0.35 |
| >10   | 56 | 2.1 | 62.8 | 0.40 | 60 | 2.1 | 36.5 | 0.30 | 43 | 1.7 | 11.9 | 0.40 |
| >Mean | 23 | 2.3 | 46.9 | 0.42 | 18 | 2.1 | 56.6 | 0.38 | | | | |

SYRIA

|       | | | | | | | | | | | |
|-------|---|---|---|---|---|---|---|---|---|---|---|
| All   | 75 | 1.8 | 10.5 | 0.18 | 131 | 2.1 | 22.9 | 0.20 | 159 | 1.9 | 13.0 | 0.16 |
| >10   | 42 | 1.8 | 10.5 | 0.21 | 69 | 2.1 | 22.9 | 0.22 | 73 | 1.9 | 13.0 | 0.25 |
| >Mean | 10 | 2.0 | 71.1 | 0.89 | 25 | 2.1 | 37.9 | 0.46 | 25 | 2.0 | 37.6 | 0.39 |

TAIWAN_PROVINCE OF CHINA

|       | | | | | | | | | | | |
|-------|---|---|---|---|---|---|---|---|---|---|---|
| All   | 32 | 1.5 | 4.7 | 0.36 | 20 | 1.5 | 14.2 | 0.34 | 18 | 1.6 | 51.1 | 0.37 |
| >10   | 17 | 1.7 | 15.8 | N/A | 15 | 1.5 | 14.2 | N/A | 12 | 1.6 | 51.1 | 0.46 |

TAJIKISTAN

|       | | | | | | | | | | | |
|-------|---|---|---|---|---|---|---|---|---|---|---|
| All   | 26 | 2.4 | 18.4 | 0.68 | 16 | 2.2 | 11.6 | N/A | 7 | 1.7 | 5.9 | N/A |
| >10   | 16 | 2.4 | 18.4 | N/A | 10 | 2.2 | 11.6 | N/A | | | | |

TANZANIA

|       | | | | | | | | | | | |
|-------|---|---|---|---|---|---|---|---|---|---|---|
| All   | 11 | 2.6 | 11.1 | N/A | 10 | 2.0 | 8.6 | N/A | 12 | 2.2 | 12.0 | N/A |
| >10   | 9 | 2.6 | 11.1 | N/A | 7 | 2.1 | 11.1 | N/A | 8 | 2.2 | 12.0 | N/A |

THAILAND

|       | | | | | | | | | | | |
|-------|---|---|---|---|---|---|---|---|---|---|---|
| All   | 127 | 2.8 | 37.4 | 0.55 | 164 | 2.1 | 18.4 | 0.47 | 208 | 2.2 | 40.9 | 0.24 |
| >10   | 68 | 2.8 | 37.4 | N/A | 97 | 2.1 | 18.4 | 0.28 | 117 | 2.2 | 40.9 | 0.20 |
| >Mean | 16 | 2.9 | 52.5 | N/A | 21 | 2.4 | 58.5 | N/A | 21 | 2.1 | 74.5 | 0.61 |

TRINIDAD AND TOBAGO

| | | | | | | | | | | | | |
|---|---|---|---|---|---|---|---|---|---|---|---|---|
| All | 9 | 1.8 | 7.6 | 0.91 | 9 | 1.9 | 13.5 | N/A | 8 | 1.5 | 4.2 | N/A |

TUNISIA

| | | | | | | | | | | | | |
|---|---|---|---|---|---|---|---|---|---|---|---|---|
| All | 81 | 2.1 | 10.6 | 0.29 | 80 | 2.2 | 15.2 | 0.33 | 101 | 2.0 | 14.0 | 0.37 |
| >10 | 46 | 2.1 | 10.6 | 0.25 | 49 | 2.2 | 15.2 | 0.29 | 49 | 2.0 | 14.0 | 0.25 |
| >Mean | 12 | 2.0 | 37.7 | N/A | 12 | 2.1 | 43.3 | 0.64 | 16 | 2.0 | 37.5 | 0.48 |

TURKEY

| | | | | | | | | | | | | |
|---|---|---|---|---|---|---|---|---|---|---|---|---|
| All | 232 | 2.4 | 27.1 | 0.34 | 221 | 2.1 | 16.4 | 0.17 | 380 | 2.0 | 19.5 | 0.16 |
| >10 | 128 | 2.4 | 27.1 | 0.29 | 128 | 2.1 | 16.4 | 0.19 | 208 | 2.0 | 19.5 | 0.16 |
| >Mean | 51 | 2.5 | 33.7 | 0.32 | 41 | 2.3 | 39.6 | 0.69 | 66 | 2.0 | 43.4 | 0.20 |

TURKMENISTAN

| | | | | | | | | | | | | |
|---|---|---|---|---|---|---|---|---|---|---|---|---|
| All | 47 | 2.1 | 13.6 | 0.34 | 48 | 2.7 | 49.6 | 0.80 | 70 | 1.8 | 7.2 | 0.17 |
| >10 | 28 | 2.1 | 13.6 | 0.36 | 31 | 2.7 | 49.6 | 0.70 | 38 | 1.8 | 10.1 | 0.24 |
| >Mean | 12 | 2.3 | 30.1 | N/A | 14 | 2.7 | 49.6 | 0.61 | 13 | 2.2 | 54.6 | N/A |

UKRAINE

| | | | | | | | | | | | | |
|---|---|---|---|---|---|---|---|---|---|---|---|---|
| All | 509 | 2.0 | 21.2 | 0.17 | 183 | 2.3 | 39.8 | 0.34 | 187 | 2.4 | 75.0 | 0.44 |
| >10 | 277 | 2.0 | 21.2 | 0.13 | 111 | 2.3 | 39.8 | 0.32 | 109 | 2.4 | 75.0 | 0.49 |
| >Mean | 104 | 2.1 | 34.1 | 0.22 | 48 | 2.3 | 39.8 | 0.35 | 39 | 2.0 | 51.7 | 0.54 |

UNITED ARAB EMIRATES

| | | | | | | | | | | | | |
|---|---|---|---|---|---|---|---|---|---|---|---|---|
| All | 41 | 1.6 | 10.3 | 0.33 | 34 | 1.7 | 10.1 | 0.22 | 38 | 1.6 | 4.7 | 0.35 |
| >10 | 29 | 1.6 | 10.3 | 0.28 | 25 | 1.7 | 10.1 | 0.33 | 21 | 1.7 | 89.2 | 0.48 |

UNITED KINGDOM

| | | | | | | | | | | | | |
|---|---|---|---|---|---|---|---|---|---|---|---|---|
| All | 460 | 1.9 | 13.5 | 0.07 | 382 | 2.0 | 39.9 | 0.15 | 402 | 2.0 | 37.6 | 0.19 |
| >10 | 296 | 1.9 | 13.5 | 0.07 | 252 | 2.0 | 39.9 | 0.14 | 250 | 2.0 | 37.6 | 0.14 |
| >Mean | 43 | 2.0 | 108.5 | 0.21 | 39 | 2.0 | 116.7 | 0.16 | 33 | 1.9 | 117.5 | 0.22 |

UNITED STATES

| | | | | | | | | | | | | |
|---|---|---|---|---|---|---|---|---|---|---|---|---|
| All | 6523 | 1.9 | 23.9 | 0.02 | 5390 | 1.9 | 26.7 | 0.03 | 5160 | 1.9 | 26.6 | 0.03 |
| >10 | 3846 | 1.9 | 23.9 | 0.03 | 3377 | 1.9 | 26.7 | 0.03 | 3131 | 1.9 | 26.6 | 0.03 |
| >Mean | 703 | 1.9 | 72.0 | 0.04 | 616 | 1.9 | 80.9 | 0.04 | 543 | 1.9 | 89.9 | 0.06 |

URUGUAY

| | | | | | | | | | | | | |
|---|---|---|---|---|---|---|---|---|---|---|---|---|
| All | 30 | 2.1 | 9.9 | 0.41 | 37 | 2.3 | 15.1 | 0.63 | 35 | 2.0 | 19.1 | 0.46 |
| >10 | 18 | 2.1 | 10.8 | 0.49 | 29 | 2.3 | 15.1 | 0.56 | 25 | 2.0 | 19.1 | 0.41 |

UZBEKISTAN

| | | | | | | | | | | | | |
|---|---|---|---|---|---|---|---|---|---|---|---|---|
| All | 176 | 2.0 | 14.2 | 0.13 | 129 | 2.3 | 34.7 | 0.41 | 120 | 2.0 | 18.9 | 0.29 |
| >10 | 107 | 2.0 | 14.2 | 0.13 | 69 | 2.3 | 34.7 | 0.38 | 64 | 2.0 | 18.9 | 0.25 |
| >Mean | 32 | 2.1 | 39.5 | 0.65 | 32 | 2.3 | 34.7 | 0.65 | 24 | 2.1 | 35.0 | 0.97 |

VENEZUELA

| | | | | | | | | | | | | |
|---|---|---|---|---|---|---|---|---|---|---|---|---|
| All | 209 | 1.8 | 16.0 | 0.10 | 205 | 1.8 | 14.4 | 0.10 | 196 | 1.8 | 11.8 | 0.15 |
| >10 | 130 | 1.8 | 16.0 | 0.10 | 136 | 1.8 | 14.4 | 0.11 | 129 | 1.8 | 11.8 | 0.10 |
| >Mean | 29 | 1.8 | 62.6 | 0.48 | 31 | 1.9 | 70.4 | 0.30 | 32 | 1.9 | 74.4 | 0.26 |

VIETNAM

| | | | | | | | | | | | | |
|---|---|---|---|---|---|---|---|---|---|---|---|---|
| All | 13 | 1.8 | 6.7 | N/A | 70 | 2.3 | 44.8 | 0.30 | 94 | 2.2 | 36.4 | 0.58 |
| >10 | 7 | 1.9 | 12.0 | N/A | 38 | 2.3 | 44.8 | 0.36 | 63 | 2.2 | 36.4 | 0.31 |
| >Mean | | | | | 12 | 2.4 | 44.7 | 1.00 | 14 | 2.1 | 69.7 | N/A |

YEMEN

| | | | | | | | | | | | | |
|---|---|---|---|---|---|---|---|---|---|---|---|---|
| All | 33 | 2.6 | 34.0 | 0.78 | 38 | 2.1 | 8.3 | 0.67 | 58 | 1.9 | 13.3 | 0.55 |
| >10 | 18 | 2.6 | 34.0 | 1.00 | 19 | 2.1 | 10.8 | N/A | 38 | 1.9 | 13.3 | N/A |
| >Mean | 7 | 2.6 | 34.0 | N/A | 8 | 2.1 | 22.5 | N/A | 13 | 2.1 | 42.6 | N/A |

YUGOSLAVIA

| | | | | | | | | | | | | |
|---|---|---|---|---|---|---|---|---|---|---|---|---|
| All | 73 | 3.0 | 24.5 | N/A | 75 | 2.4 | 14.6 | 0.35 | 136 | 2.1 | 12.6 | 0.29 |
| >10 | 42 | 3.0 | 24.5 | N/A | 49 | 2.4 | 14.6 | 0.38 | 81 | 2.1 | 12.6 | 0.33 |
| >Mean | 19 | 2.7 | 21.7 | N/A | 19 | 2.6 | 28.6 | N/A | 29 | 2.3 | 30.8 | 0.69 |

ZAMBIA

| | | | | | | | | | | | | |
|---|---|---|---|---|---|---|---|---|---|---|---|---|
| All | 10 | 1.7 | 7.5 | 0.48 | 16 | 2.0 | 14.0 | 0.49 | 24 | 1.9 | 8.2 | 0.58 |
| >10 | 8 | 1.7 | 10.1 | N/A | 10 | 2.0 | 14.0 | N/A | 15 | 1.9 | 10.1 | N/A |

ZIMBABWE

| | | | | | | | | | | | | |
|---|---|---|---|---|---|---|---|---|---|---|---|---|
| All | 16 | 1.9 | 11.4 | 0.56 | 19 | 1.6 | 3.3 | 0.27 | 9 | 1.6 | 3.3 | 0.39 |
| >10 | 12 | 1.9 | 11.4 | | 9 | 2.1 | 38.1 | | | | | |